\documentclass[aps,prx,reprint,amsmath,amssymb,floatfix,superscriptaddress]{revtex4-1}
\usepackage[dvipsnames]{xcolor}
\usepackage{graphicx}
\usepackage{dcolumn}
\usepackage[T1]{fontenc} 
\usepackage{bm}
\usepackage{caption}
\usepackage{subcaption}
\usepackage{amsfonts}
\usepackage{verbatim}
\usepackage[caption=false]{subfig}
\usepackage{dsfont}

\newcommand{\beq}{\begin{equation}}
\newcommand{\eeq}{\end{equation}}
\newcommand{\beqa}{\begin{eqnarray}}
\newcommand{\eeqa}{\end{eqnarray}}
\newcommand{\braket}[1]{\left<#1\right>}

\begin{document}


\title{Quantum Altermagnetic Instability in Disordered Metals}

\author{Alberto Cortijo}
\email{alberto.cortijo@csic.es}
 \affiliation{Instituto de Ciencia de Materiales de Madrid (ICMM), Consejo Superior de Investigaciones Científicas (CSIC), Sor Juana Inés de la Cruz 3, 28049 Madrid, Spain}

\date{\today}
             
\begin{abstract}
The possibility of a zero-temperature altermagnetic instability in anisotropic two-dimensional electron systems in the diffusive regime is analyzed in the presence and absence of spin--orbit coupling. Allowing for ferromagnetism, a phase diagram is built as a function of the parameter that controls anisotropy and the strength of the interactions. It is found that at zero spin--orbit coupling ferromagnetism only dominates at small values of the anisotropy and the coupling constant. Larger values of these parameters favour the formation of altermagnetism. At finite spin--orbit coupling, a paramagnetic phase competes with the other two, and a quantum critical point appears. The phase transition from the paramagnetic to the magnetically ordered phases is of second order, while the phase transition between ferromagnet and altermagnet states is first order. The altermagnetic phase is not sensitive to small magnetic fields, coexisting with a field-induced magnetization.
\end{abstract}

\maketitle


\section{\label{sec:intro}Introduction}

Altermagnets (AM) have recently emerged as a new class of collinear antiferromagnets in which electronic bands can be spin split even in the absence of spin--orbit coupling and without generating any net magnetization\cite{sinova22,song25}. This spin--projected band degeneracy lifting has motivated a rapidly expanding range of proposed device concepts. On the spintronics side, AM have been advanced as platforms for spin-dependent transport and electrically driven spin manipulation without the constraints associated with ferromagnetic stray fields\cite{Bai23,zhang24}, and as potential platforms for novel multiferroic functionalities\cite{aoyama24,wu24,zhu24}. In parallel, AM/superconductor heterostructures have been proposed as hybrid systems in which spin degeneracy in superconductors can be lifted by proximity effect to an AM without the need for external magnetic fields\cite{Schaffer25,Jiang2025,Alipourzadeh25}, with implications towards topological superconductivity\cite{ghorashi24,Hadjipaschalis25}. Across these diverse directions, a common and practical requirement stands out: the altermagnetic order parameter must reliably \emph{form} and remain stable in materials and samples under realistic conditions. Also, it is important for these applications that no ferromagnetic magnetization appears, as reported in $\text{CrSb}$\cite{Zhou25} or bulk $\text{MnTe}$\cite{Kluczyk2024}. From a microscopic standpoint, this requirement is nontrivial because altermagnetism is not only a statement about a staggered magnetization (antiferromagnetism), but also about a staggered \emph{anisotropy} encoded in the crystal environment of the two sublattices (as dictated by its magnetic space-group symmetry, recalled below). Without both ingredients, the characteristic lifting of Kramers degeneracy does not materialize, and the band-structure signatures that underpin many proposed functionalities are absent. Moreover, the most experimentally relevant settings for applications---thin films, interfaces, and heterostructures---are intrinsically affected by disorder\cite{zeng25,Chakraborty25,palacios25}, spatial inhomogeneity\cite{ornellas25}, strain\cite{Chakraborty24}, and diffusive electron dynamics\cite{Zarzuela25,Sun25}. This motivates the development of microscopic mechanisms for the \emph{formation} of altermagnetic order that remain operative beyond idealized clean limits\cite{Vasiakin25}.
\begin{figure}[b]
\centering
\includegraphics[width=0.4\textwidth]{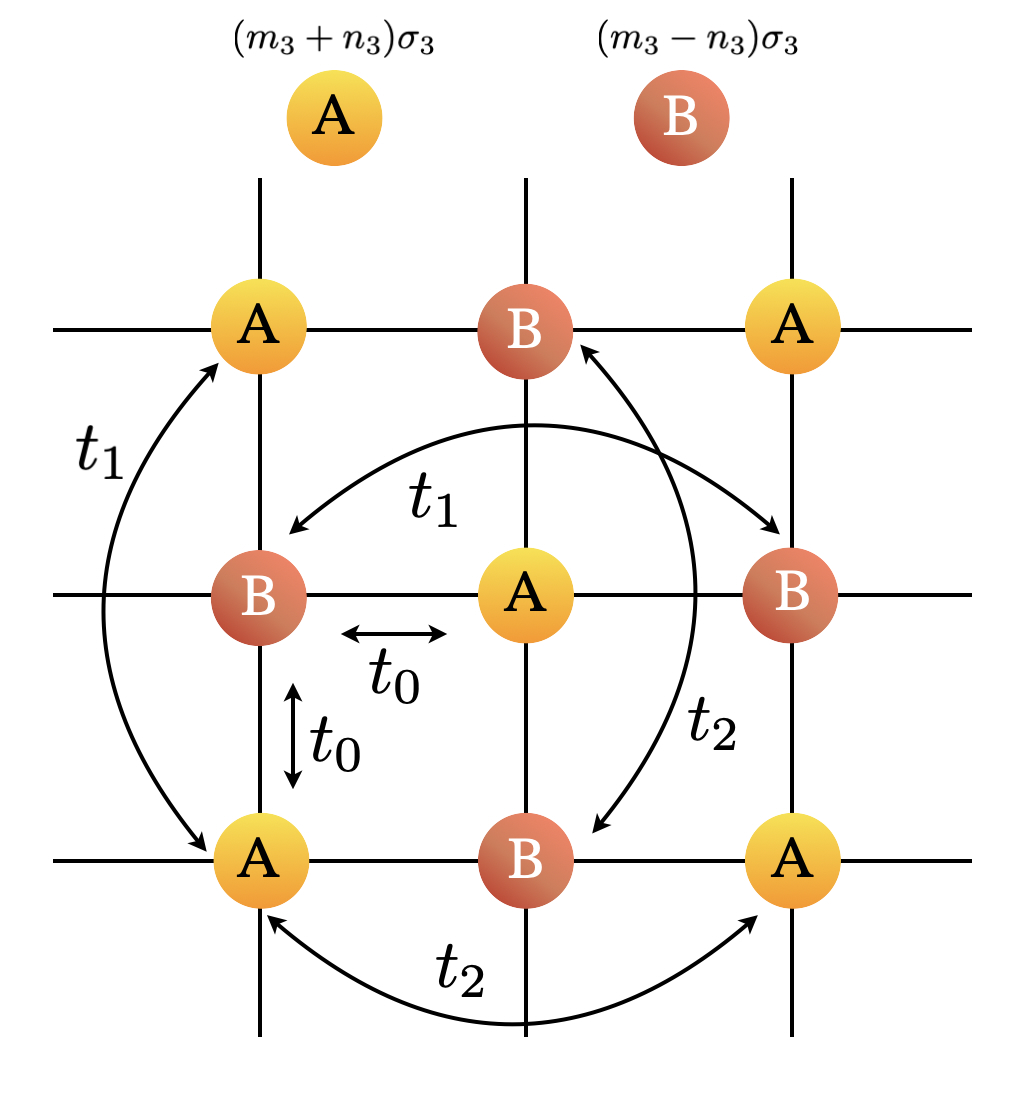}
\caption{\label{fig:lattice}(Color online) Illustration of the hopping structure of the anisotropic model considered in the text. The hopping amplitudes $r_1$ and $r_2$ break inversion symmetry. These amplitudes  together with the indicated magnetic moments  are invariant under $\{C_n\Vert C_2^{S}\}$. When spin--orbit coupling is included, the symmetry group stabilizing the altermagnetic phase is $\text{C}_4\cdot\text{T}$.}
\end{figure}

In this work we analyze the interaction-driven itinerant magnetism in a diffusive electron system as a route to the stabilization of altermagnetic phases in realistic setups. In diffusive metals, magnetic fluctuations couple efficiently to hydrodynamic spin diffusion modes, producing magnetically ordered ground states in the free energy\cite{Nayak2003}. While in isotropic settings these diffusive interaction effects are known to enhance itinerant ferromagnetism in two dimensions, the central question we address is whether the same diffusive route to magnetism can instead favor altermagnetic order once the symmetry-enforced staggered anisotropy is present. To make this symmetry constraint explicit, we briefly recall the defining space-group structure of AM. Altermagnets are characterized by a nontrivial operation
$\{C_n\Vert C_2^{S}\}$ (spin-space groups), which combines a real-space crystal rotation $C_n$ detached from a $\pi$ rotation (or spin flip) in spin space $C_2^{S}$. Crucially, the real-space part implies that the two magnetic (crystallographic) sublattices are related by a \emph{rotation} rather than by inversion or translation. Microscopically, this forces the local electronic environment of one sublattice to be a rotated copy of the other, so that the kinetic energy inherits a \emph{staggered anisotropy} (e.g., rotated hopping amplitudes) between sublattices. This staggered anisotropy need not make the Fermi surface strongly anisotropic; rather, its key role is revealed once magnetic order develops:
The staggered magnetization combines with the staggered anisotropy to lift the Kramers degeneracy and produce spin-projected bands whose anisotropy is directly proportional to both the staggered anisotropy and the altermagnetic order parameter.
The important message is that the altermagnetic instability may appear even if the system does not develop an altermagnetic instability in the clean limit. The key ingredient is the presence of staggered anisotropy in the diffusive regime, both in the presence and in the absence of spin--orbit coupling. If, together with an antiferromagnetic order parameter, we allow for a ferromagnetic component, we can construct the phase diagram as a function of the staggered-anisotropy parameter and the interaction strength. The inclusion of a Rashba spin--orbit term changes the symmetry considerations in the system, because the stabilizing symmetry is no longer dictated by $\{C_n\Vert C_2^{S}\}$ but by the magnetic group $\text{C}_4\cdot\text{T}$, where $\text{T}$ is the time reversal operation. For d--wave altermagnets, the difference is not essential in practice, as the d--wave altermagnetic order parameter is an even function of the momentum components. We will find that the altermagnetic phase competes and is the leading magnetic phase in large portions of the phase diagram. We also show that the altermagnetic phase is not sensitive to small perpendicular magnetic fields.
\section{\label{sec:model}Model for a staggered anisotropic system}

We will use the following simple tight-binding model featuring the ingredients discussed in the previous section that allow for the AM ordering\cite{Maier2023,Roig24}: 

\begin{eqnarray}\label{eq:kineticH}
H=&-&t_0\sum_{\bm k}\psi^+_{\bm k}(\cos ak_1+\cos ak_2)\tau_1\psi_{\bm k}\\\nonumber
&-&\underbrace{(t_1+t_2)}_{t}\sum_{\bm k}\psi^+_{\bm k}(\cos2ak_1+\cos2ak_2)\tau_0\psi_{\bm k}\\\nonumber
&-&\underbrace{(t_1-t_2)}_{\Delta}\sum_{\bm k}\psi^+_{\bm k}(\cos2ak_1-\cos2ak_2)\tau_3\psi_{\bm k}.
\end{eqnarray}
The hopping amplitudes displaying the staggered anisotropy are depicted in Fig.~\ref{fig:lattice}. The Pauli matrices $\bm{\tau}$ represent the sublattice degree of freedom, $\psi^+_{\bm k \sigma}=(c^+_{A\bm k\sigma},c^+_{B\bm k\sigma})$. We will also consider as a model Hamiltonian for magnetic interactions an on-site Hubbard model, parametrized by the interaction strength $U$, where $a=\pm1$ labels the two sublattices. Within a mean field approach we can write,

\begin{eqnarray}\label{Hmeanfield}
H_{int}=\sum_{i,a=\pm}\bm s_{i,}\,\cdot\,\hat{\bm S}_{i,a}
-\frac{3}{8U}\sum_{i,a=\pm} \hat{\bm S}_{i,a}\,\cdot\, \hat{\bm S}_{i,a}.
\end{eqnarray}
As usual, $\bm s_i=\frac{1}{2}c^{+}_{i,\sigma}\bm \sigma\, c_{i,\sigma'}$ represents the electronic spin, and $\hat{\bm S}_a$ represents the Hubbard--Stratonovich (HS) on-site magnetization\cite{coleman2015}. For simplicity, we will assume the HS fields $\hat{\bm S}_a$ to condense along the $\hat{\bm z}$ direction, $\hat{\bm S}_{a}\to\braket{\bm S_{a}}=S_{a,3}\,\hat{\bm z}$. It is also convenient to parametrize the condensate $S_{3,a}$ as $S_{3}\to m_3+ n_3$ for the sublattice A ($\tau=1$) and $S_{3}\to m_3- n_3$ for sublattice B ($\tau=-1$) as depicted in Fig.~\ref{fig:lattice}.

Further assuming the HS fields to be homogeneous, we can write $H_{\text{int}}$ in momentum space,
\begin{eqnarray}\label{Hmeanfield2}
H_{int}&\simeq&\sum_{\bm{k}}m_{3}\,c^{+}_{\bm k\sigma}\tau_0\sigma_{3} c_{\bm k\sigma'}+n_{3}\,
c^{+}_{\bm k\sigma}\tau_3\sigma_{3} c_{\bm k\sigma'}\\\nonumber
&-&\frac{1}{2U}\sum_{\bm q}(m^2_3+n^2_3),
\end{eqnarray}
after a trivial redefinition of $U$.
The derivation of the effective $\bm{k}\cdot\bm{p}$ model for the kinetic part of the \emph{bonding} states at the $\Gamma$ point, can be found in Appendix \ref{app:kpmodel}. The resulting effective Hamiltonian is of course invariant under the operation $\{C_{n}\Vert C_{2}^{S}\}$. Although altermagnetism does not require the presence of spin--orbit coupling to lift the Kramers degeneracy, it will be convenient to incorporate into the discussion a Rashba spin--orbit term, that is invariant under the magnetic group $\text{C}_4\cdot\text{T}$, but not under $\{C_{n}\Vert C_{2}^{S}\}$ (see the discussion at the end of Sec. \ref{sec:intro}). To complete the effective model we will analyze, we need to incorporate the spin--orbit (SOC) coupling. Although it can be done starting from the initial four-band model, it is enough to add it to the continuum two-band model in the form of the term in the Hamiltonian $H_{\text{SO}}=\sum_{\bm k}\alpha_R\, b^{+}_{\bm k}(\bm \sigma\times\bm k)\cdot\hat{z}\,b_{\bm k}$. So after adding a scalar disorder potential $V_{\bm k,\bm k'}$ in momentum space\cite{maiani25}, our model reads,

\begin{eqnarray}\label{Hamiltonian}
H&=&\sum_{\bm k}b^{+}_{\bm k \sigma}\,[\varepsilon_{\bm k}\sigma_0+\bm h(\bm{k})\cdot\bm{\sigma}]\,b_{\bm k \sigma'}\\\nonumber
&+&\sum_{\bm k,\bm k+\bm q}V_{\bm k\bm q}b^{+}_{\bm k \sigma}b_{\bm k+\bm q \sigma}-\frac{1}{2U}\sum_{\bm q}(m^2_3+n^2_3),
\end{eqnarray}
where we have defined $\varepsilon_{\bm k}=\bm k^2/2m_a$, and the vector
\begin{equation}
  \bm h(\bm{k})
  =
  \bigl(\alpha_R k_2,\,-\alpha_R k_1,\,
  m_3 + \delta n_3 a^2(k_1^2 - k_2^2)
  \bigr).
  \label{eq:h_model}
\end{equation}
We have defined the dimensionless parameter $\delta=4\Delta/t_0$ that controls the degree of staggered anisotropy, and $a$ is the lattice spacing. This is the standard Hamiltonian of a two-dimensional electron system in the presence of Rashba spin--orbit coupling, magnetization, and an AM contribution to the Zeeman term. A comment is in order here. We have kept quadratic terms of $\bm k$ in (\ref{eq:h_model}), so, to this order, the anisotropy only enters in the term $\delta\, n_3\, a^2 (k^2_1-k^2_2)$. This is a feature of the expansion around the $\Gamma$ point. Expanding around the $X$ and $Y$ points in the Brillouin Zone, the anisotropy parameter $\delta$ would enter directly at quadratic order instead. Also, it is important to keep in mind that in the 4-band model, the Néel order parameter is a local operator (i.e. independent of momentum) while it becomes momentum-dependent \emph{after} the reduction to the 2-band model.

\section{\label{sec:BSequation} Nonlinear $\bm \sigma$-model approach to disorder}

\subsection{\label{subsec:NLsigmaM} Effective Field Theory}
In the problem of disordered electron systems, it is apparent that the low energy, long wavelength excitations that transport charge, spin and/or heat in the diffusive regime are no longer electronic Bloch states, but spin/charge diffusion modes. For this reason, it was natural as well that in the problem of interacting disordered electron systems, charge/spin diffusion modes are the natural hydrodynamic degrees of freedom that would couple to the effective magnetization at a mean field level\cite{kirkpatrick96,Chamon2000,Nayak2003,Belitz2016}. The dynamics of the diffusion modes in a ferromagnetic two-dimensional electron gas with a Rashba spin--orbit term has been extensively studied in the literature. The fundamental difference with the problem discussed here is that, in the Hamiltonian (\ref{Hamiltonian}), the AM order parameter can be interpreted as a nematic Zeeman term close to the Fermi level:
$\varepsilon_{\text{Zeeman}}=m_3+\delta a^2k^2_F n_3\cos2\varphi$. This angular dependence will induce fundamental changes in the diffuson propagators.

Following the standard procedure of the replica-based nonlinear $\sigma$ model (NL$\sigma$M) treatment of disorder\footnote{We will limit ourselves to outline this method following references~\cite{kirkpatrick00,Chamon2000,Nayak2003}. We recommend to the reader to visit these references.}, we introduce $\alpha=1,...,N$ replicas of the action associated to the Hamiltonian (\ref{Hamiltonian}), \emph{construct} the advanced/retarded spaces, and construct the replicated partition function $\braket{Z^N}_{\text{dis}}=\braket{\int\Pi_\alpha\mathcal{D}b^{+}_\alpha\,b_\alpha \mathcal{D}(m_{3}, n_{3})e^{i\sum_\alpha\mathcal{S}_\alpha}}_{\text{dis}}$, where $\braket{\ldots}_{\text{dis}}$ is the Gaussian integration over the disorder field $V$, with zero mean $\braket{V_{\bm k}}=0$, and isotropic variance $\braket{V_{\bm k}\,V_{\bm k'}}=\Gamma_0 \delta(\bm k+\bm k')$ ($\Gamma_0$ represents the scattering probability). We will assume that the potential $V$ belongs to the AI symmetry class in the Altland-Zirnbauer classification, meaning that the disorder potential is invariant under spin rotations and time reversal operation\cite{Altland97}. The Gaussian integration over the disorder potential $V_{\bm k}$ with this variance leads to a quartic interaction for fermions that mixes replica indices, 
\begin{eqnarray}\label{scattprob}
H_{\text{dis}}=\Gamma_0\sum_{\bm k\bm k'\bm q}\sum^{N}_{\alpha,\beta=1}(b^{+}_{\bm{k},\alpha}b_{\bm{k}-\bm{q},\alpha})(b^{+}_{\bm{k}',\beta}b_{\bm{k}'+\bm{q},\beta}).
\end{eqnarray}
Thermodynamic and (Kubo-like) transport physical properties are then constructed from retarded and advanced correlation functions due to causality constraints. For this reason we will extend the replica space to an advanced/retarded space. In free systems, this implies adding to the otherwise real poles of Green and correlation functions a positive/negative small imaginary part $\pm i\eta$.

The next step is to perform a Hubbard-Stratonovich transformation to convert the quartic interaction (\ref{scattprob}) into the coupling of a given fermionic bilinear with a replica matrix field $Q_{\alpha\beta}\sim b^+_{\alpha}b_{\beta}$. The final step is to perform the functional integration over the fermionic variables, to obtain the effective action, at zero temperature, 
\begin{eqnarray}\label{replicaaction}
\braket{Z^N}_{\text{dis}}=\int\,\mathcal{D}Q_{\alpha\beta}\mathcal{D}(m_{3},n_{3}) e^{-i\mathcal{S}[Q_{\alpha\beta},m_{3},n_{3}]},
\end{eqnarray}
with 
\begin{eqnarray}\label{replicaaction2}
\mathcal{S}&=&-\text{Tr}\ln\left[(G^{R/A}_{\text{free}})^{-1}\delta_{\alpha\beta}+\frac{i}{2\tau}Q^{R/A}_{\alpha\beta}\right]\\\nonumber
&+&\text{Tr}(Q^{R/A}_{\alpha\beta})^2+\frac{1}{2U}\sum_{\bm q}((m_{3}\delta^{R/A}_{\alpha\beta})^2+(n_{3}\delta^{R/A}_{\alpha\beta})^2),
\end{eqnarray}
where $\text{Tr}$ stands for summation over retarded/advanced, replica, spin and momentum-frequency summations. ${G}^{R/A}_{\text{free}}$ is the free, \emph{clean} fermionic Green function in the retarded/advanced space. The scattering time is now related to the total density of states and $\Gamma_0$ by the expression $\tau=1/2\pi\nu\Gamma_0$. The replica limit, $\ln Z_{\text{eff}}=\lim_{N\to0}\partial_{N}\braket{Z^N}_{\text{dis}}$, is taken at the end of the procedure.

At this stage it is also standard to consider first the saddle-point equation for the matrix fields $Q_{\alpha\beta}$, which is equivalent to the self-consistent Born approximation (SCBA) for the fermionic self-energy in other treatments of the disorder. The solution of the saddle-point equation $\delta \mathcal{S}/\delta Q=0$ is, making explicit the spin, retarded/advanced and replica indices, and treating $m_3$ and $n_3$ as background fields,
\begin{equation}\label{saddlepoint}
Q^{R/A}_{\alpha\beta,\mu\nu}=\delta_{\alpha\beta}\delta_{\mu\nu}\begin{pmatrix}
1 & 0\\
0 & -1
\end{pmatrix}_{R/A}.
\end{equation}
The solution (\ref{saddlepoint}) implies that the replica symmetry is not broken, although there is still a notion of symmetry breaking between advanced and retarded replicas, since there is now a finite quantity $\pm i/2\tau$ that replaces the small parameter $i\eta$. Besides this technical point, what matters to us is that the nonzero saddle-point matrix $Q^{R/A}_{\alpha\beta,\mu\nu}$ represents a finite relaxation time for the electronic degrees of freedom and satisfies the constraints $\text{Tr}[Q^{R/A}_{\alpha\beta,\mu\nu}]=0$ (due to the structure in the retarded/advanced space) and $(Q^{R/A}_{\alpha\beta,\mu\nu})^2=\mathbf{1}$.

The low energy degrees of freedom in the resulting theory are the fluctuations of the matrix $Q$ around the saddle-point solution (\ref{saddlepoint}) that respect the previous constraints. Following \cite{Efetov80}, we can parametrize the saddle-point manifold as 
\begin{equation}
\tilde{Q}^{R/A}_{\alpha\beta,\mu\nu}=e^{-\frac{W}{2}}Q^{R/A}_{\alpha\beta,\mu\nu} e^{\frac{W}{2}},
\end{equation}
where $\tilde{Q}^{R/A}_{\alpha\beta,\mu\nu}$ satisfies the same constraints as $Q^{R/A}_{\alpha\beta,\mu\nu}$ if the fields $W$ satisfy the properties $\{ W,Q\}=0$ and $\text{Tr}[W]=0$. These fluctuation fields $W$ can be written in the retarded/advanced space as,
\begin{eqnarray}
W=\begin{pmatrix}
0 & B \\
-B^{+} & 0 
\end{pmatrix}_{R/A},
\end{eqnarray}
and the fields $B$ can be written in turn in the spin space as $B=S_0\,\sigma_0+\sigma_i\,S_i$. Let us remember that the fields $S_0$ and $S_i$ are still matrices in the replica space ($(\alpha\,\beta)$ indices) and in spin space ($(\mu,\,\nu)$ indices). These fields will be the low-energy, long-wavelength fields that represent the charge and spin diffusons in the diffusive limit. Expanding Eq.~(\ref{replicaaction2}) in powers of $S_0$ and $S_i$ up to second order, we proceed to construct the effective theory of the hydrodynamic modes that couple to the fields $m_3$ and $n_3$. The next section is devoted to the computation of the effective propagators $D_{ij}(\omega,\bm q)$ for the spin diffusons $S_i$, using the Bethe-Salpeter formalism\cite{altland06}.

As a final comment, we will assume that the replica symmetry holds, so all diffusons and fields, viewed as matrices in the replica space, are diagonal, so they scale with $N$, and it is straightforward to take the replica limit $N\to0$.


\subsection{\label{subsec:BSformalism}Bethe Salpeter equation for the diffuson propagator}

The fermionic Green functions in the SCBA and in retarded/advanced space are
\begin{equation}
  G^{R/A}(\varepsilon,\bm k)
=
\frac{1}{\varepsilon - \varepsilon_{\bm k}
- \bm h(\bm k)\cdot\bm\sigma
\pm i/(2\tau)},
  \label{eq:Greenfunction}
\end{equation}
with $\varepsilon_{\bm k}$ and $\bm h(\bm k)$ defined after Eq.~(\ref{Hamiltonian}). In what follows, we will use $\mu=0$ and $\mu=(1,2,3)$ for the charge and spin modes, respectively. The retarded/advanced Green functions in Eq.~(\ref{eq:Greenfunction}) tell us that the Bloch states of the free system are no longer the low energy degrees of freedom. This role will be taken by the charge and spin diffusons $S_0$ and $S_i$.

To build the dynamics of these diffuson modes, we need to compute the correlation functions that describe such dynamics through the Bethe Salpeter equation (BS) formalism\cite{altland06}. The basic building blocks  of the BS equation for spin diffusons are the  bare spin--resolved particle/hole polarization function $\Pi_{ij}$ constructed from Eq.~(\ref{eq:Greenfunction}) ($\mathrm{Tr}$ stands for trace over spin matrices and $d^2\bm k=\nu \,d 
\varphi\, d\varepsilon_{\bm k}$):
\begin{widetext}
\begin{eqnarray}
  \Pi_{ij}(\omega,\bm{q})
  =\nu
  \int\,{d\varphi}\int d\varepsilon_{\bm{k}}
  \int d\varepsilon\, 
  \mathrm{Tr}\Big[
    \sigma_i\,
    G^R\!\big(\varepsilon+\tfrac{\omega}{2},\bm{k}+\tfrac{\bm{q}}{2}\big)\,
    \sigma_j\,
    G^A\!\big(\varepsilon-\tfrac{\omega}{2},\bm{k}-\tfrac{\bm{q}}{2}\big)
  \Big],
  \label{eq:app_Pi_def}
\end{eqnarray}
and the spin--resolved diffuson propagator $D_{ij}$ that satisfies the following BS equation in the ladder approximation (repeated indices imply summation):
\begin{eqnarray}
D_{ij}(\omega,\bm{q})
  =
  \Pi_{ij}(\omega,\bm{q})
  +
  \Gamma_0
  \Pi_{il}(\omega,\bm{q})\,D_{lj}(\omega,\bm{q}).
  \label{eq:BS}
\end{eqnarray}
\end{widetext}
Eq.~(\ref{eq:BS}) can be cast into the form of a kinetic equation:
\begin{eqnarray}
\mathcal{L}_{il}D_{lj}\equiv\left[\delta_{il}-\Gamma_0\Pi_{il}\right]D_{lj}=\Pi_{ij}.
\end{eqnarray}
The zero modes of the operator $\mathcal{L}_{ij}$ will thus define the propagating spin diffuson modes.

The details of the computation of the polarization function $\Pi_{ij}$ can be found in Appendix \ref{app:polarization} in the diffusive regime and in the Dyakonov-Perel (DP) regime. With the expressions (\ref{eq:hydropole2}), (\ref{eq:finalorder12}), and (\ref{averagedorder2}) we can write the kinetic equations for the spin diffusons $(\bm S_{\perp},S_3)$:
\begin{eqnarray}\label{spindiffusons}
[i\omega+Dq^2+\Gamma(n_3)]\bm S_{\perp}-2\,\bm m\times \bm S_{\perp}=0,
\end{eqnarray}
\begin{eqnarray}\label{S3diffuson}
[i\omega+Dq^2+2\Gamma_{\text{DP}}(0)]S_3=0,
\end{eqnarray}
where we have defined the magnetization precession $\bm m=(0,0,m_3)$, and the \emph{altermagnetic} Dyakonov-Perel (DP) relaxation time $\Gamma(n_3)\equiv\Gamma_{DP}+\lambda\tau n^2_3=\tau\alpha_R^2k^2_F+\lambda\tau n^2_3$ (recall that $\lambda=\delta^2 a^4k^4_F$). From now on, we will use the notation $\Gamma_{\text{DP}}=\tau\alpha_R^2k^2_F$. 

The reason why the AM order parameter $n_3$ enters in the spin diffuson dynamics as a spin relaxation term for the perpendicular components of the spin diffuson, similar to the effect of the Rashba term, can be understood as follows: The AM order parameter acts as a momentum-dependent Zeeman term (as can be seen in Eq.~(\ref{eq:h_model})). The in-plane spin of an  electron with momentum $\bm {k}$ will experience a precessional motion given by the effective magnetic field $\lambda n_3\cos2\varphi$ along the $\hat{z}$ direction. After being scattered to another momentum $\bm k'$, the new precessional motion of the in-plane spin will change as the effective magnetic field has changed. After many random scattering events that change momentum, the spin motion has experienced a randomization of the AM effective magnetic field, with zero mean value, but with finite variance, that leads to an effective relaxation rate\cite{Zarzuela25,Sun25}. The charge diffuson $S_0$ and the third component of the spin diffuson, $S_3$, are not modified by the presence of the FM and AM order parameters, so we will not consider them in the next discussions.

The physical scenario now changes with respect to the one for magnetic instabilities in clean electronic systems. In clean systems, the FM and AM order parameters enter at the mean-field level as constant and momentum-dependent Zeeman terms, respectively\cite{Maier2023,Roig24,Chen25}. In the present scenario, however, they enter in fundamentally different ways: FM order enters as a spin-precession term, while the AM order parameter enters through the spin-relaxation time. This implies that, as we will see, their effect on the effective potential will be different.

Having identified the hydrodynamic degrees of freedom that couple to the order parameters $m_3$ and $n_3$\cite{Castellani84}, we can write down the effective potential for the in-plane spin diffuson modes $S_i$ ($i,j=(1,2)$) after normalizing by the system size and characteristic time ($V_{\text{eff}}=\frac{S_{\text{eff}}}{L^2 T}$),
\begin{eqnarray}\label{effectiveactiondiffusons}
V_{\text{eff}}=\frac{1}{2U}(m^2_3+n^2_3)-\int\frac{d\omega}{2\pi}\frac{d^2\bm q}{4\pi^2}\,S^{*}_i\mathcal{L}_{ij}S_j,
\end{eqnarray}

\subsection{\label{subsec:BSformalism2}Effective action for the order parameters}

We proceed now to integrate out the diffuson modes in Eq.~(\ref{effectiveactiondiffusons}) in order to write an effective action only in terms of the fields $m_3$ and $n_3$. First of all, it is convenient to change the spin diffusons to the chiral basis, $S_{\pm}=(S_1\pm i S_2)/\sqrt{2}$, so the kernel of $\mathcal{L}_{ij}$ becomes diagonal:
\begin{eqnarray}\label{effectiveactionchiral}
V_{\text{eff}}&=&\frac{1}{2U}(m^2_3+n^2_3)-\sum_{r=\pm}\int\frac{d\omega}{2\pi}\frac{d^2\bm q}{4\pi^2}\cdot\\\nonumber
&&\cdot\,S^{*}_{r}\left(i\omega+D q^2+\Gamma(n_3)+2i\,r\,m_3\right)S_{r}.
\end{eqnarray}

In the limit $\alpha_R=0$ and $n_3= 0$, the action (\ref{effectiveactionchiral}) coincides with the effective action computed in Ref.\cite{Nayak2003}. The action $S_{\text{diff}}$ provides a simple understanding of the different effects of the ferromagnetic (FM) order parameter $m_3$ and the AM one $n_3$. While $m_3$ acts as degeneracy breaking between the two chiral diffusons, $n_3$ enters as a contribution to the Dyakonov-Perel spin relaxation time. In the small $m_3$ limit, this degeneracy breaking term will enter as a natural infrared cutoff in the integral over $\omega$ when $\alpha_R=n_3=0$, leading to the non-analytic quadratic term $m^2_3\ln(m_3)$ in the effective action which drives the FM instability for any $U$\cite{Chamon2000,Nayak2003}. The presence in this scenario of a $\Gamma_{\text{DP}}$ will strongly modify this infrared structure, leading to an analytic quadratic term $\sim m^2_3\log \Gamma_{\text{DP}}$, forcing the transition to appear at a critical $U_c$. In what follows we analyze how this scenario changes with the presence of an AM order parameter, and how this order parameter $m_3$ competes with $n_3$.

Integrating out the diffusons $S_{\pm}$ we arrive to the \emph{trace-log}, one-loop form of the effective action for the $m_3$ and $n_3$ degrees of freedom:

\begin{eqnarray}\label{tracelogaction}
V_{\text{eff}}&=&\frac{1}{2U}(m^2_3+n^2_3)-\sum_{r=\pm}\int\frac{d\omega}{2\pi}\int\frac{d^2\bm q}{4\pi^2}\cdot\\\nonumber
&&\cdot\,\ln\left[i\omega+Dq^2+\Gamma(n_3)+2i\,r\,m_3\right].
\end{eqnarray}
The expansion of the action (\ref{tracelogaction}) in powers of small $m_3$ and $n_3$ is legitimate when $\alpha_R>0$. However, we want to discuss the limit $\alpha_R=0$, so we will perform the integration over $\omega$ and $\bm q$ in Eq.~(\ref{tracelogaction}). It is convenient to Wick rotate the frequencies to the imaginary axis in the expression (\ref{tracelogaction}). Introducing dimensionless variables and parameters $s=D q^2/\Lambda$, $m=m_3/\Lambda$, $\rho_R=\Gamma_{\text{DP}}/\Lambda$, and $g=U/\pi^2D$ (recall that $\lambda$ is already dimensionless and $\Lambda=1/\tau$), and multiplying the previous expression by $4\pi^2D$, we can write
\begin{eqnarray}\label{integralaction}
V_{\text{eff}}&=&\frac{2}{g}(m^2+n^2)\\\nonumber
&-&\int^{2}_{0} ds\,L(s)\,\ln\Big[\big(s+\rho_R+\lambda n^2\big)^2+4m^2\Big].
\end{eqnarray}
This is the integral representation of the effective potential that depends on three free parameters, $(\rho_R, g,\lambda)$. The details of the transformation to this representation of the effective potential can be found in Appendix \ref{app:mathdetailsintegral}. Specifically, the function $L(s)$ can be found in Eq.~(\ref{Lfunction}).

\begin{figure*}[t]
\centering
  \includegraphics[width=\textwidth]{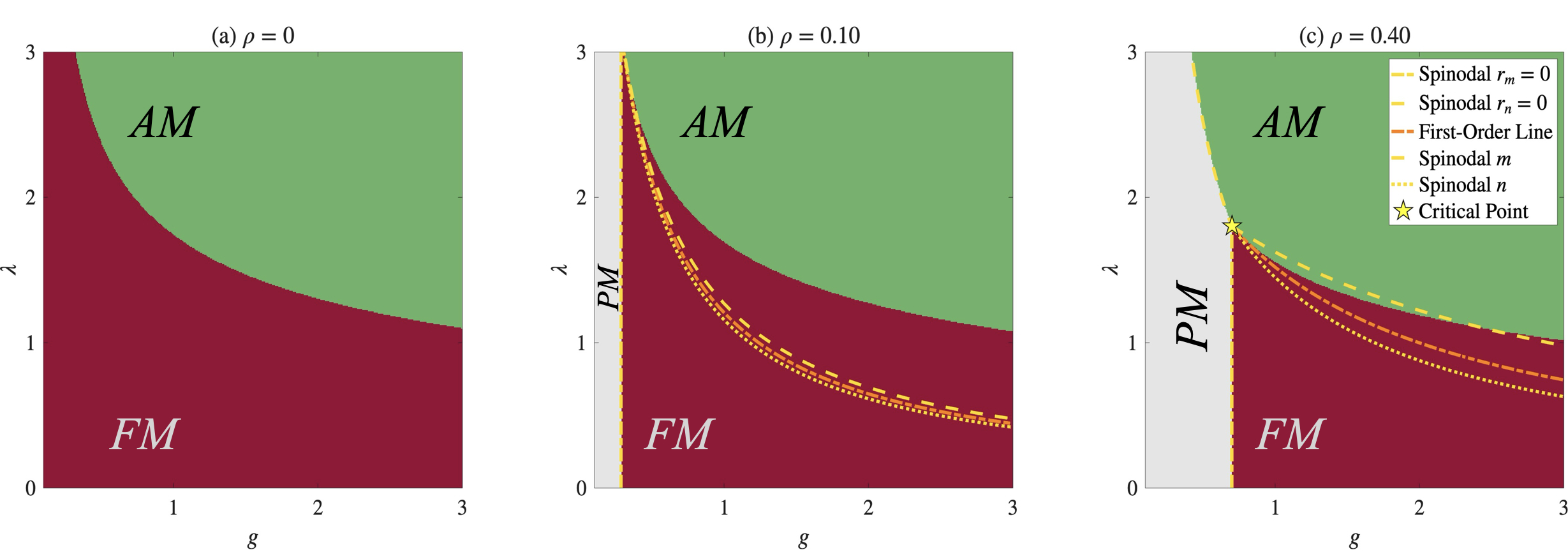} 
  \caption{(Color online) Phase diagram corresponding to the effective potential in Eq.~(\ref{integralaction}) for three different values of the dimensionless Rashba coupling $\rho_R$. (a) $\rho_R=0$. Due to the nonanalytic nature of the effective potential when $\rho_R=0$, the Landau expansion in Eq.~(\ref{potentialapp}) fails at the qualitative level. However, Fig.~\ref{fig:potential_evolution} shows that the phase transition between the FM (ferromagnetic) and AM (altermagnetic) phases is of first order. (b), (c) $\rho_R=0.1$ and $\rho_R=0.4$. At finite $\rho_R$, a PM (paramagnetic) phase appears [the point $(0,0)$ becomes a stable minimum for some values of $g$ and $\lambda$]. The dotted, dash-dotted (orange), and dashed lines within the FM phase correspond to the two spinodal lines and the FM--AM phase-boundary line, respectively. The agreement between the analytical and numerical FM--AM phase-boundary lines is qualitative at best. The solid yellow lines represent the PM--FM and PM--AM boundary lines. As these transitions are second order [implying truly small values of the $m$ and $n$ order parameters], the agreement between numerics and analytics is complete. These two lines become spinodal lines upon entering the FM and AM regions. In panel (b), the critical point lies outside the plotted region of the phase diagram.}
  \label{fig:threepanels}
\end{figure*}

\section{\label{sec:phasediagram}Phase diagram}
In this section we analyze the phase diagram for a diffusive two-dimensional anisotropic electron system that emerges from the effective potential (\ref{integralaction}). We will do so in terms of the interaction strength $g$ and the anisotropy parameter $\lambda$, for several representative values of the dimensionless Rashba coupling $\rho_R$.

\subsection{\label{subsec:ferrocase}Ferromagnetic instability for $\rho_R=\lambda=0$}

As a check, we can analyze the action $V_{\text{eff}}$ in  in the limit $\Gamma_{\text{DP}}=\lambda=0$, that is, the FM instability in a diffusive $2D$ isotropic electron system. In this limit, the quadratic-logarithmic term dominates over the quartic and higher order terms (that are suppressed by powers of the inverse cutoff $\Lambda$) and we get\cite{Finkelstein82,Chamon2000,Nayak2003},

\begin{eqnarray}\label{eq:ferrocase}
V_{\text{eff}}[m]&\simeq&\frac{2}{g}m^2-\int^{1}_{0} ds\,s\,\ln\Big[s^2+4m^2\Big]\\\nonumber
&\simeq&\frac{2}{g}m^2-2m^2\ln m^2.
\end{eqnarray}

Now it is easy to see that the potential  in (\ref{eq:ferrocase}) develops a global FM minimum away from $m=0$ for any value of the interaction strength $g$\cite{andreev98,Chamon2000,Nayak2003}:
\begin{equation}
2m\approx e^{-\frac{1}{2g}},
\end{equation}
and the origin $m=0$ is always a maximum with infinite negative curvature.

\subsection{\label{subsec:AFcase}Altermagnetic instability for $\rho_R=0$}

In the previous section, we rederived the nonanalytic quadratic term that gives rise to ferromagnetism in two-dimensional diffusive systems. This $m^2\log m^2$ term makes the ferromagnetic instability unavoidable, but the order parameter in Eq.~(\ref{eq:ferrocase}) might be too small to be detected experimentally for small values of $g$. Let us now see whether nonanalyticities also appear in the theory of altermagnetism.

Setting $m=0$ and $\rho_R=0$ in Eq.~(\ref{integralaction}) and expanding for small values of $n$, we obtain
\begin{eqnarray}\label{eq:AMcase}
V_{\text{eff}}[n]&\simeq&\left(\frac{2}{g}-4\lambda\log2\right)n^2\\\nonumber
&+&\frac{1}{2}\lambda^2\left[3-2\log(\lambda n^2)\right]n^4.
\end{eqnarray}
The nonanalytic structure of the potential in Eq.~(\ref{eq:AMcase}) appears in the quartic term, just simply because the order parameter $n$ enters with a different power than $m$ in (\ref{integralaction}). The main consequence is that the quadratic term, that controls the phase transition is analytic, so the phase transition takes place when $g=g_c=\frac{1}{\lambda\,2\log2}$, and the nonanalytic quartic term is positive for small values of $n$. Contrary to the ferromagnetic case, there is a Stoner criterion for the altermagnetic instability to occur that depends on the anisotropy parameter $\lambda$.

\subsection{\label{subsec:generalcase}Phase diagram for finite $\rho_R$}
Let us describe the thermodynamics of the system when we allow for both order parameters $m$ and $n$ for finite values of $\rho_R$. Now, the effective potential $V_{\text{eff}}[n,m]$ does not display nonanalytic terms, so it is illustrative to perform an expansion up to quartic powers of $n$ and $m$:
\begin{eqnarray}\label{potentialapp}
V_{\text{eff}}[m,n]&=&r_m m^2+u_m m^4+r_n n^2+u_{n}n^4\\\nonumber
&+&u_{mn}m^2n^2.
\end{eqnarray}
To understand qualitatively the phase diagram for finite $\rho_R$ (the interested reader might find more information in the references \cite{Liu73,Kosterlitz76}), we will perform the analysis in terms of the Landau parameters $(r_m,r_n)$, and $(u_m,u_n,u_{mn})$. The expressions of the Landau coefficients in terms of $\rho_R$, $g$, and $\lambda$ will be displayed later.

The important element to pay attention to is the presence of a term that couples $m^2$ and $n^2$. Also, we can check that all the coefficients $u$ are strictly positive, which implies global thermodynamic stability and that the cross coupling term $m^2 n^2$ is an energy penalty to form ground states with simultaneous nonzero values of $m$ and $n$, ruling out coexistent phases.

Let us discuss the stability of the extremal points of the effective potential $V_{\text{eff}}[m,n]$ in terms of the expression (\ref{potentialapp}). For that, we need the stationary equations,
\begin{subequations}
\begin{eqnarray}
2m(r_m+2u_m m^2+u_{mn}n^2)=0,\label{gapeqm}
\\
2n(r_n+2u_n n^2+u_{mn}m^2)=0.\label{gapeqn}
\end{eqnarray}
\end{subequations}
and the expression for the Hessian matrix $H$ associated to $V_{\text{eff}}[m,n]$:
\begin{subequations}
\begin{eqnarray}
\frac{\partial^2 V_{\text{eff}}}{\partial m^2}&=&2(r_m+6 u_m m^2+u_{mn}n^2),
\\
\frac{\partial^2 V_{\text{eff}}}{\partial n^2}&=&2(r_n+6 u_n n^2+u_{mn}m^2),
\\
\frac{\partial^2 V_{\text{eff}}}{\partial m\,\partial n}&=&4 u_{mn} m\, n.
\end{eqnarray}
\end{subequations}
The expression for the Hessian determinant is thus, 
\begin{eqnarray}
\det(H)&=&4r_m\, r_n+4m^2(r_m u_{mn}+6r_n u_m)\\\nonumber 
&+&4n^2(r_n u_{mn}+6 r_m u_n)+24u_{mn}(u_m m^4+u_n n^4)\\\nonumber
&+&12m^2 n^2(12u_m u_n -u^2_{mn}).
\end{eqnarray}

The stationary equations give four extremal points: $(0,0)$, $(m_0,0)$, $(0,n_0)$, and $(m_{\star},n_{\star})$, with $m_0=\sqrt{-\frac{r_m}{2u_m}}$, $n_0=\sqrt{-\frac{r_n}{2u_n}}$, and $m_\star$ and $n_\star$ defined later. From here it is easy to see that the solutions exist for $r_m<0$ and $r_n<0$, respectively. However, they do not require $r_m<0$ and $r_n<0$ simultaneously.

The point $(0,0)$ gives the following value for the Hessian determinant, $\det(H)=4r_m r_n$. This is a stable minimum only when $r_m>0$ and $r_n>0$ simultaneously. This implies that the point $(0,0)$ ceases to be a minimum at $r_m=0$, and $r_n=0$, being these conditions the two first spinodal lines. For $r_m>0$ and $r_n>0$, the point $(0,0)$ is a maximum.

\begin{figure}   
\centering
    \includegraphics[width=\linewidth]{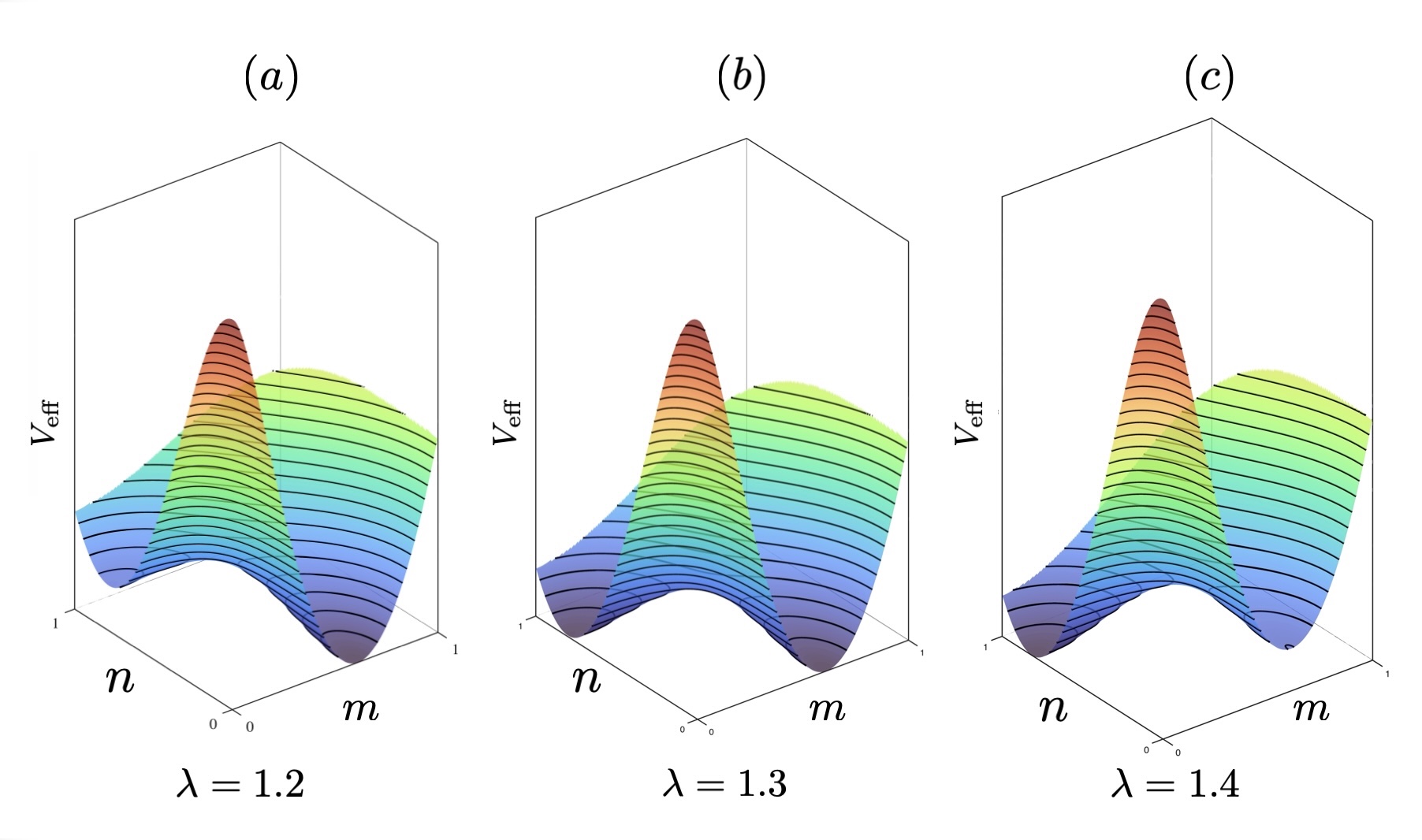}
    \caption{(Color online) Plot of the effective potential Eq.~(\ref{integralaction}) for $\rho_R=0$, $g=2$, and three different values of $\lambda$ showing the first order phase transition scenario: Several minima at finite values of the order parameters $m$ and $n$ compete across the boundary line. In (a) the AM minimum is local and FM is the global one. In (b) the two minima are degenerate, and (c) show that the FM becomes local while AM is the global minimum.}
\label{fig:potential_evolution}
\end{figure}

For the point $(m_0,0)$ to be a minimum, the stationary equation Eq.~(\ref{gapeqm}) requires $r_m<0$ (substituting the value of $m_0$ in Eq.~(\ref{gapeqm}) gives automatically $n=0$ without requiring $r_n>0$). The determinant for this solution is $\det(H)=-4r_m\left[2r_n-r_m\frac{u_{mn}}{u_m}\right]$, and the prefactor $-4r_m$ is strictly positive. Therefore, the local stability of the $(m_0,0)$ phase relies on the term inside brackets, and the change of sign of $r_n$ does not imply that this point loses stability, that occurs when $r_n=r_m\frac{u_{mn}}{2u_m}$. Also, as discussed before, the solution $(m_0,0)$ ceases to be a stable minimum at $r_m=0$. Interestingly, if $r_n>0$, it is easy to see directly from the effective potential that $V_{\text{eff}}[m_0,n]$ is a strictly growing function of $n$, so $V_{\text{eff}}[m_0,n]>V_{\text{eff}}[m_0,0]$, implying that for $r_n>0$, $(m_0,0)$ is a \emph{global} minimum. When $r_n<0$, we can be above or below the spinodal line defined above. Summarizing, the extremal point $(m_0,0)$ is a minimum between the two spinodal lines $r_m=0$ and $r_n=r_m\frac{u_{mn}}{2u_m}$.

The stability analysis of the point $(0,n_0)$ parallels the discussion of $(m_0,0)$, just exchanging the roles between $r_m$ and $r_n$ and $u_m$ and $u_n$. In this case, the spinodal line lies in the line $r_m=r_n\frac{u_{mn}}{2u_n}$, and this point is a stable minimum between the spinodal lines $r_n=0$ and $r_m=r_n\frac{u_{mn}}{2u_n}$.

Let us analyze the stability of  the last extremal point $(m_\star,n_\star)$. The simultaneously nonvanishing solutions of the stationary equations are
\begin{subequations}
\begin{eqnarray}
m_\star&=&\sqrt{\frac{u_{mn}r_n-2u_n r_m}{2(4u_m u_n-u^2_{mn})}},
\\
n_\star&=&\sqrt{\frac{u_{mn}r_m-2u_m r_n}{2(4u_m u_n-u^2_{mn})}}.
\end{eqnarray}
\end{subequations}
The Hessian determinant is, after particularizing it for this point, $\det(H)=16m^2_\star\,n^2_\star (4u_m\,u_n-u^2_{mn})$. From this expression we can see that, if $u^2_{mn}<4u_m\,u_n$, the determinant is positive, so $(m_\star,n_\star)$ is a stable local minimum and there is a potential phase where there is coexistence of phases (both $m$ and $n$ are simultaneously nonzero). If $u^2_{mn}>4u_m\,u_n$, the determinant becomes negative, and the point $(m_\star,n_\star)$ is a saddle point, implying a first order phase transition between the stationary points $(m_0,0)$ and $(0,n_0)$ points. The best way to discuss this first order phase transition line is to realize that it takes place when the value of the effective potential is the same for the two points $(m_0,0)$ and $(0,n_0)$, $V_{\text{eff}}[m_0,0]=V_{\text{eff}}[0,n_0]\rightarrow u_mr^2_n=u_nr^2_m$. Since these solutions imply $r_n<0$ and $r_m<0$, we can write, for the first order line,
\begin{equation}
r_n=r_m\sqrt{\frac{u_n}{u_m}}.
\end{equation}
Interestingly, the condition $u^2_{mn}>4u_m u_n$ that, together with $r_m<0$ and $r_n<0$, guarantees that the following hierarchy of inequalities hold:
\begin{equation}
\frac{u_{mn}}{2u_m}>\sqrt{\frac{u_n}{u_m}}>\frac{2u_n}{u_{mn}}.\label{spinodalregion}
\end{equation}

Let us translate all this abstract analysis to our problem. To do that, we need the expressions of the coefficients in Eq.~(\ref{potentialapp}) as functions of the parameters $\rho_R$, $g$, and $\lambda$. They are easy to obtain after the expansion in powers of $n$ and $m$ of $V_{\text{eff}}[n,m]$. We list them below: 
\begin{subequations}
\begin{eqnarray}
r_m&=&\frac{2}{g}-4F_1(\rho_R),\label{eq:rm}
\\
r_n&=&\frac{2}{g}-2\lambda F_2(\rho_R), \label{eq:rn}
\\
u_m&=&\frac{8}{3}\frac{2+6\rho_R+3\rho^2}{\rho^2(1+\rho_R)^2(2+\rho_R)^2}, \label{eq:um}
\\
u_n&=&\lambda^2 F_1(\rho_R),\label{eq:un}
\\
u_{mn}&=&\frac{8\lambda}{\rho_R(1+\rho_R)(2+\rho_R)},\label{eq:umn}
\end{eqnarray}
\end{subequations}
where we have defined the auxiliary functions
\begin{subequations}
\begin{eqnarray}
F_1(\rho_R)&=&\log\left(\frac{(1+\rho_R)^2}{\rho_R(2+\rho_R)}\right),
\\
F_2(\rho_R)&=&\rho_R\log\rho_R-2(1+\rho_R)\log(1+\rho_R)\\\nonumber
&+&(2+\rho_R)\log(2+\rho_R).
\end{eqnarray}
\end{subequations}
With these equations, we can write down the expressions for the spinodal ( and phase boundary) lines. The lines describing the stability of the paramagnetic (PM) phase towards the ferromagnetic (FM) and altermagnetic (AM) phases are:
\begin{subequations}
\begin{eqnarray}
(r_m=0)&\leftrightarrow& g_{c,m}=\frac{1}{2F_1(\rho_R)},\label{criticalferroU}
\\
(r_n=0)&\leftrightarrow&  g_{c,n}=\frac{1}{\lambda F_2(\rho_R)}.\label{criticalAM}
\end{eqnarray}
\end{subequations}
The condition (\ref{criticalferroU}) implies a Stoner-like criterion for the FM transition to occur, $g>g_{c,m}$. Also, this critical line is independent of $\lambda$. The equation (\ref{criticalAM}) defines the critical line between the  PM and AM phases. These lines can be seen in Fig.~\ref{fig:threepanels},b, and Fig.~\ref{fig:threepanels},c.
In the limit of small $\rho_R$, it simply gives $g_{c,n}\simeq1/(\lambda\,\ln4)$, that is, a finite value of $g_{c,n}$. This is in contrast to the isotropic FM case, where the critical $g_{c,m}$ approaches zero. It implies that, in the limit $\rho_R\sim0$ for those values of $0<g<g_{c,n}$, the only allowed instability is the FM one, as it can be seen in Fig.~\ref{fig:threepanels},a. The PM-FM and PM-AM are second order phase transitions, as the order parameters $m$ and $n$ vary continuously from zero to finite values when $r_m$ and $r_n$ move from negative to positive values. This is why in these cases, the spinodal lines coincide with the phase boundary lines.

Let us analyze now the stability of the point $(m_\star,n_\star)$ in terms of the system parameters $\rho_R$, $g$, and $\lambda$. In the previous paragraphs we have seen that the stability condition for the coexistence solution $(m_\star,n_\star)$ (to define a local minimum or a saddle point) depends on the condition $u^2_{mn}>4u_m\,u_n$. A little bit of algebra allows us to convert this condition into an expression only in terms of the dimensionless spin--orbit coupling $\rho_R$:
\begin{equation}\label{FOcondition}
\frac{6}{(2+6\rho_R+3\rho^2_R)\log\left(\frac{(1+\rho_R)^2}{\rho_R(2+\rho_R)}\right)}>1.
\end{equation}
This inequality holds for values $\rho_R$ larger than $\rho_{R,c}\approx0.035$. In the Landau theory (\ref{potentialapp}), when $\rho_R<\rho_{R,c}$, the coexistence solution $(m_\star,n_\star)$ becomes energetically favourable against a hard first order phase transition between the $(m_0,0)$ FM and $(0,n_0)$ AM phases. However, when the phase diagram is computed using the exact effective potential for $\rho_R<\rho_{R,c}$, no such coexistence phase appears. The reader is warned that this might be expected as the effective potential (\ref{potentialapp}) is just an approximate theory, and the conclusions drawn from it are qualitative at best, and for $\rho_R=0$ it completely fails, as it can be seen in Fig.~\ref{fig:threepanels},a, where a hard first order line between the FM and AM phases is observed. In Fig.~\ref{fig:potential_evolution} we have plotted the evolution of the exact effective potential $V_{\text{eff}}[m,n]$ for $\rho_R=0$ in Eq.~(\ref{integralaction}). It is seen that even at $\rho_R\to 0$, the effective potential still displays the structure of two minima located at $(m_0,0)$ and $(0,n_0)$. No trace of coexistence minimum appears. At fixed $g$, the global vs. local nature of the minima interchange when crossing the first-order boundary line determined by Eq.~(\ref{eq:FOline}): The point $(m_\star,n_\star)$ is a saddle point in the exact theory.

That said, the accuracy of this analytical expansion becomes better for larger values of $\rho_R$. In the rest of this section, we will safely stay in the parameter regime $\rho_R>\rho_{R,c}$, so the coexistence solution $(m_\star,n_\star)$ is a saddle point and the phase transition between the FM and AM phases is of first order. 

The phase-transition line between these two phases is easily computed within this Landau approach if we define the quantity $K(\rho_R)=\frac{\rho_R(1+\rho_R)(2+\rho_R)}{2}\sqrt{\frac{3F_1(\rho_R)}{2(2+6\rho_R+3\rho^2_R)}}$. Then, the first order line gets the form, written as a critical $\lambda$ as a function of $g$ (from the condition $r_n=r_m\sqrt{\frac{u_n}{u_m}}$):
\begin{equation}
\lambda_{c,FO}=\frac{1}{gF_2(\rho_R)+K(\rho_R)[1-2gF_1(\rho_R)]}.\label{eq:FOline}
\end{equation}
This first-order boundary line and the second-order boundary lines in Eqs.(\ref{criticalferroU},\ref{criticalAM}) allows us to analyze the existence of a tricritical point where these lines intersect. We can check that these lines do intersect at the following point in the $(g,\lambda)$ space:
\begin{equation}
\text{critical point}=\left(\frac{1}{2F_1(\rho_R)},\frac{2F_1(\rho_R)}{F_2(\rho_R)}\right),
\end{equation}
which is a function only of $\rho_R$.

Let us discuss the stability of the pure FM and AM solutions: when they stop to be local or global minima to be saddle points. The analysis is done as before analyzing the conditions when the Hessian determinant becomes zero when particularized for $(m_0,0)$ and $(0,n_0)$. These conditions give rise to two spinodal lines. The one describing the stability of the FM phase takes the form:
\begin{widetext}
\begin{equation}
(r_n=r_m\frac{u_{mn}}{2u_m})\leftrightarrow\lambda_{c,m}=\frac{2(2+6\rho_R+3\rho^2_R)}{3\rho_R(1+\rho_R)(2+\rho_R)+g[(2(2+6\rho_R+3\rho^2_R)F_2(\rho_R))-6\rho_R(1+\rho_R)(2+\rho_R)F_1(\rho_R)]},\label{spinodalM}
\end{equation}
while the spinodal line for the AM phase is,
\begin{equation}
(r_m=r_n\frac{u_{mn}}{2u_n})\leftrightarrow\lambda_{c,n}=\frac{4}{\rho_R(1+\rho_R)(2+\rho_R)F_1(\rho_R)+g[4F_2(\rho_R)-2\rho_R(1+\rho_R)(2+\rho_R)F^2_1(\rho_R)]}.\label{spinodalN}
\end{equation}
\end{widetext}
As described in Eq.~(\ref{spinodalregion}), the stability condition $u^2_{mn}>4u_m\,u_n$ ensures that the critical line Eq.~(\ref{eq:FOline}) appears in the phase diagram in between the two spinodal lines (\ref{spinodalM}) and (\ref{spinodalN}). In Fig.~\ref{fig:threepanels},b and Fig.~\ref{fig:threepanels},c, the exactly calculated phase diagram for $\rho_R=0.1$ and $\rho_R=0.4$ are plotted, respectively. As expected, due to the second order character of the PM-FM and PM-AM phase transitions, the spinodal lines $r_m=0$ and $r_n=0$ are accurate with the exact lines, while the phase boundary line (\ref{eq:FOline}) fails quantitatively in fitting the position of the true phase transition line, as it can be seen in Figs.\ref{fig:threepanels},b,c. In these panels we also plot the analytical expressions of the spinodal lines, Eqs.(\ref{spinodalM},\ref{spinodalN}).

As a last exercise, we can naively take the limit $\rho_R\to 0$ in all the spinodal lines. As expected, the critical value of $g$ required to produce the FM phase goes to zero, while the AM phase becomes unstable for $g<g_{c,n}\to\frac{1}{\lambda 2\log 2}$. The two spinodal lines (\ref{spinodalM}) and (\ref{spinodalN}) collapse to $g_{c,n}\to\frac{1}{\lambda 2\log 2}$ as well, while the first-order line disappears. As mentioned above, this is a shortcoming of the analytical Landau expansion, and it should not be trusted over the numerical calculations.

\begin{figure}
  \centering
  \begin{subfigure}[t]{0.42\textwidth}
    \centering
    \includegraphics[width=\linewidth]{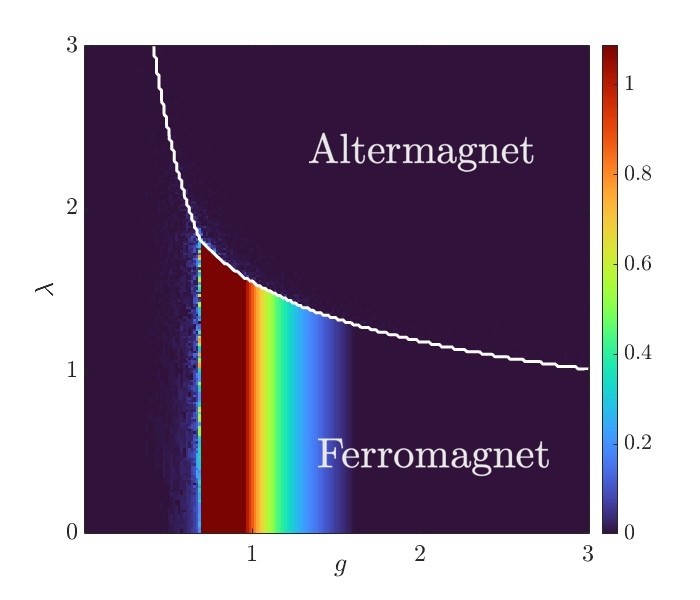}
    \caption{$\rho_R=0.4$, $b=0$.}
    \label{fig:chi0}
  \end{subfigure}\hfill
  \begin{subfigure}[t]{0.42\textwidth}
    \centering
    \includegraphics[width=\linewidth]{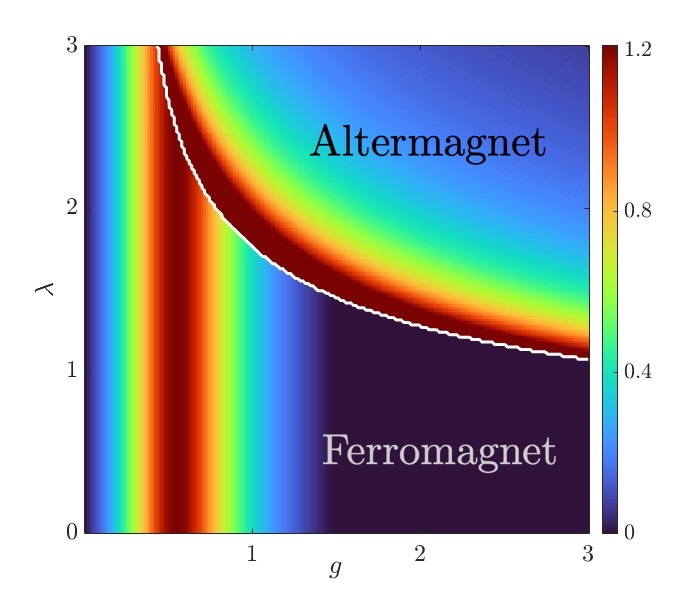}
    \caption{$\rho_R=0.4$, $b=0.05$.}
    \label{fig:chi005}
  \end{subfigure}

  \caption{(Color online) (a): Magnetic susceptibility $\chi(g,\lambda,b)$ (in arbitrary units) for $\rho_R=0.4$ and zero magnetic field. (b) $\chi(g,\lambda,b)$ for $\rho_R=0.4$ and dimensionless magnetic field $b=0.05$. The boundary lines between phases are depicted for clarity.}
  \label{fig:chi}
\end{figure}

\subsection{Effect of an external magnetic field}

\begin{figure*}\label{fig:secondorder}
\centering
\begin{subfigure}[b]{0.35\textwidth}
  \centering
  \includegraphics[width=\linewidth]{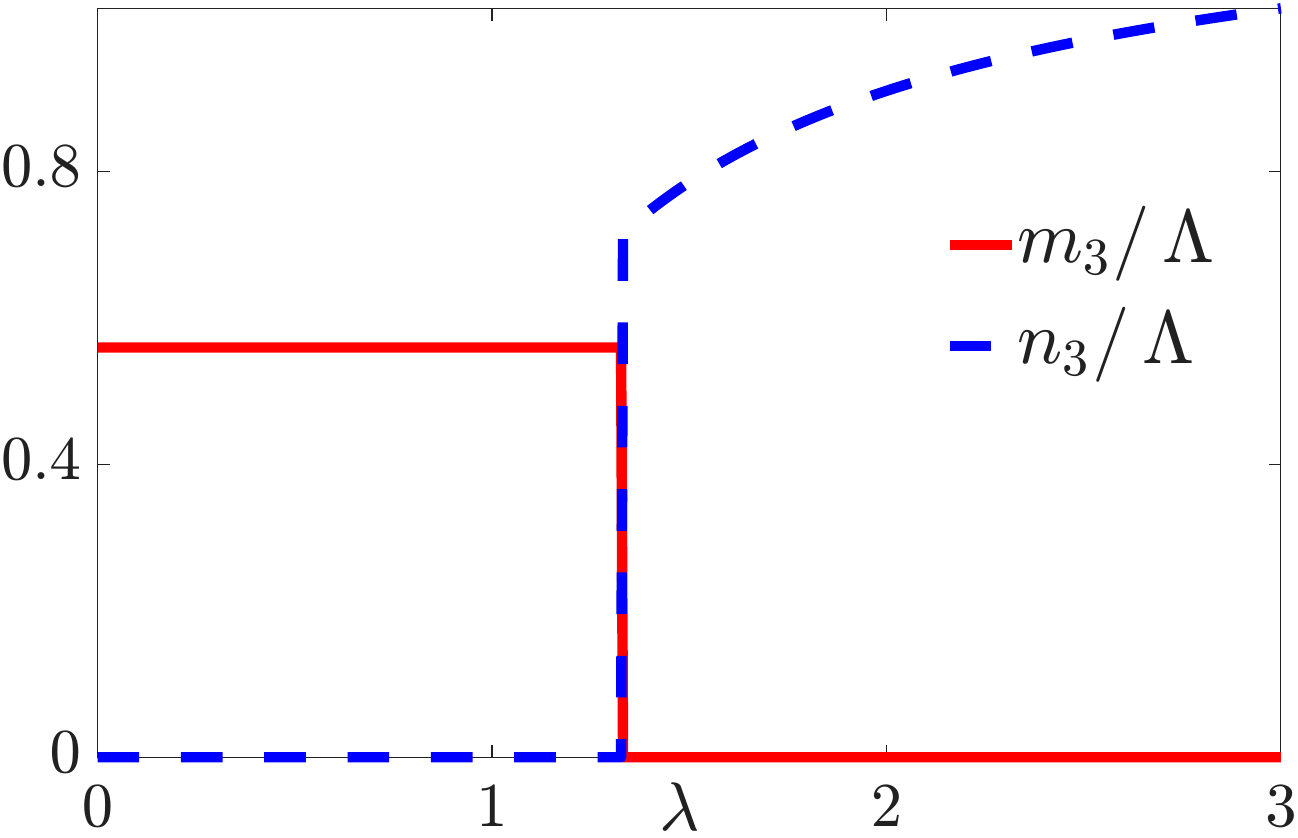}
  \caption{$\rho_R=0.4$, $g=1.5$.}
  \label{fig:gfixed}
\end{subfigure}
\begin{subfigure}[b]{0.35\textwidth}
  \centering
  \includegraphics[width=\linewidth]{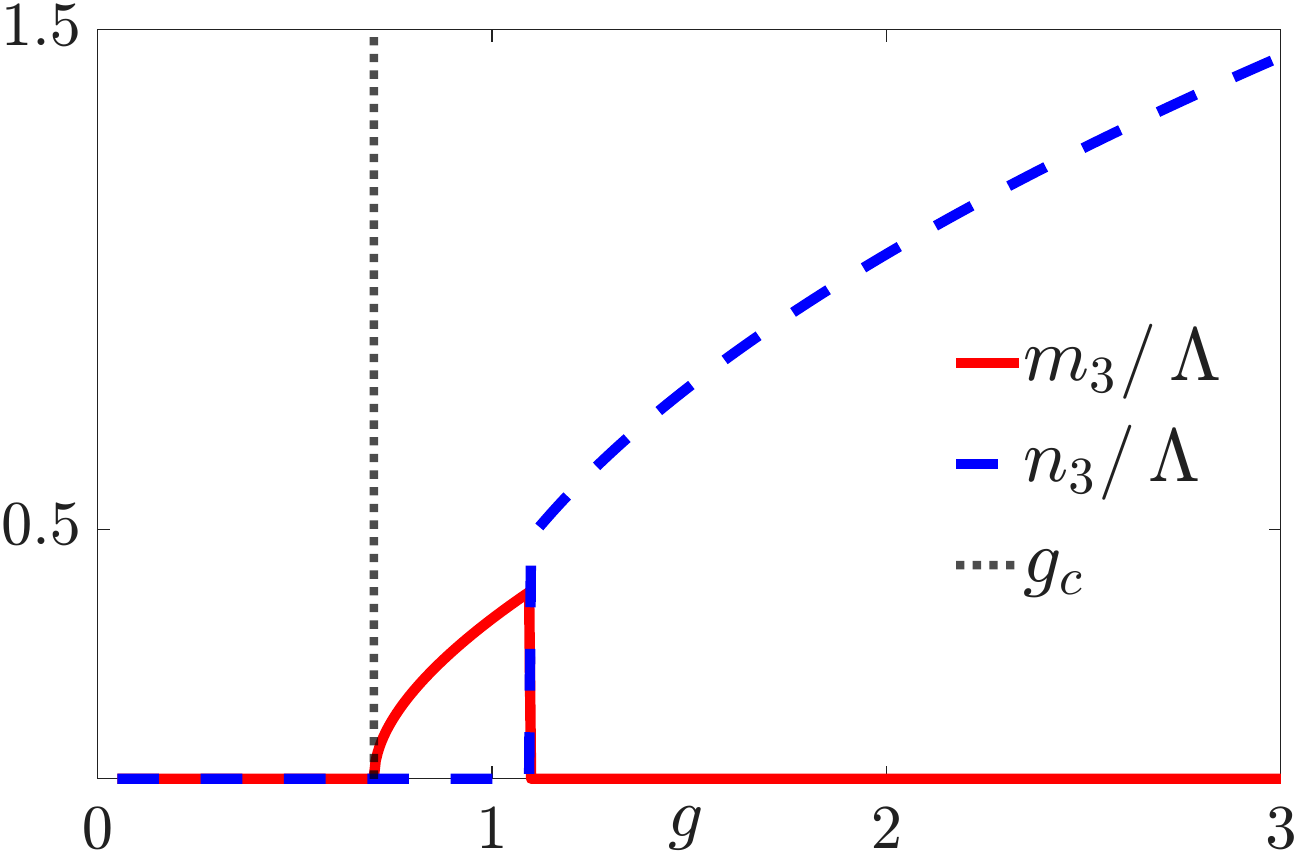}
  \caption{$\rho_R=0.4$, $\lambda=1.5$.}
  \label{fig:lambdafixed}
\end{subfigure}


\begin{subfigure}[b]{0.35\textwidth}
  \centering
  \includegraphics[width=\linewidth]{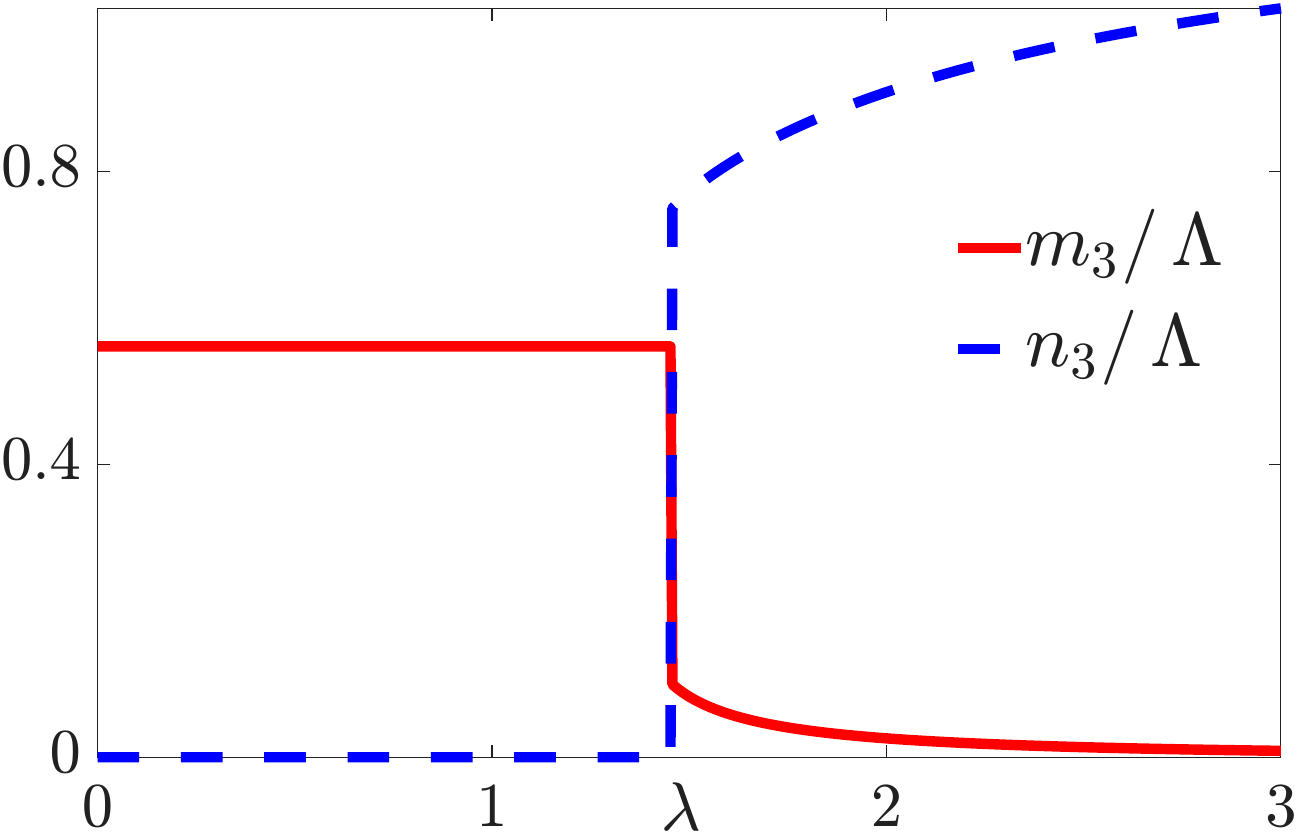}
  \caption{$\rho_R=0.4$, $g=1.5$, $b=0.05$.}
  \label{fig:gfixedB}
\end{subfigure}
\begin{subfigure}[b]{0.35\textwidth}
  \centering
  \includegraphics[width=\linewidth]{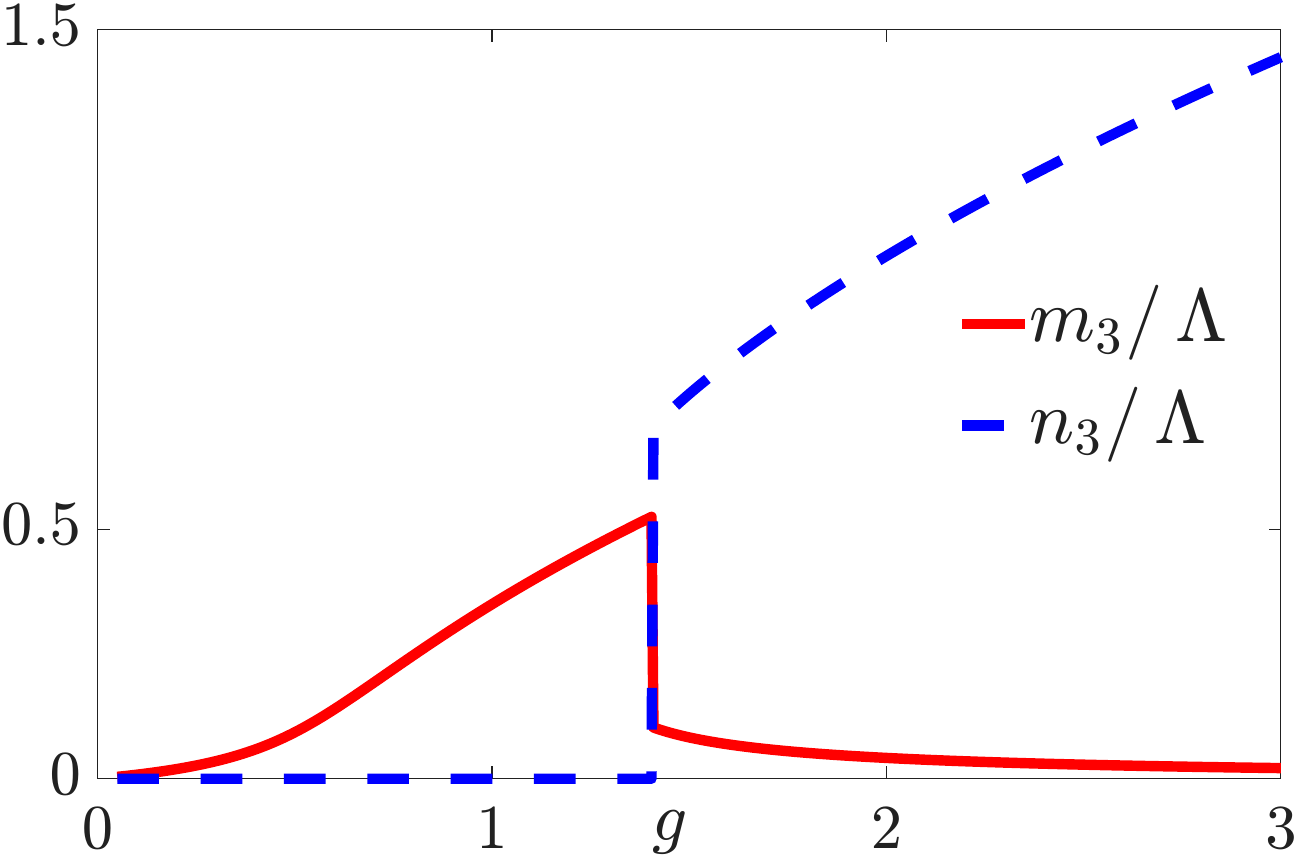}
  \caption{$\rho_R=0.4$, $\lambda=1.5$, $b=0.05$.}
  \label{fig:lambdafixedB}
\end{subfigure}

\caption{(Color online) Character of the different phase transitions. In panels (a) and (b) the order parameters $m_3/\Lambda$ and $n_3/\Lambda$ are plotted for $\rho_R=0.4$ and zero magnetic field as a function of $\lambda$ and $g$, respectively. The PM--FM transition is of second order, while the FM--AM transition is first order. Panels (c) and (d) represent the same order parameters as before, but in the presence of a finite dimensionless magnetic field $b=0.05$. The PM phase is heavily reduced, and the FM phase penetrates into the AM phase, generating phase coexistence.}
\label{fig:fourpanels}
\end{figure*}

Finally, we discuss the effect of an external magnetic field perpendicular to the system, $B_3$. Small magnetic fields enter primarily as a Zeeman term in the Hamiltonian Eq.~(\ref{Hmeanfield2}) and induce spin precession in the spin dynamics. Then, it is straightforward to include them in the dynamics of the spin diffusons (\ref{spindiffusons}) by performing the substitution $2m_3\to 2m_3+g_{\text{s}}\mu_B B_3$ (where $g_{\text{s}}$ is the effective gyromagnetic ratio, $\mu_B$ is the Bohr magneton). The dimensionless magnetic field entering in the effective potential (\ref{integralaction}) is $b=g_{\text{s}}\mu_B B_3\tau$ with the substitution $4m^2\to 4(m+b)^2$ within the logarithm in Eq.~(\ref{tracelogaction}). The magnetic susceptibility is computed numerically using the thermodynamic expression $\chi=\frac{\partial m_3}{\partial B}$ once the phase diagram is settled (see panel Fig.~\ref{fig:threepanels}(c)). In Fig.~\ref{fig:chi0} we observe that the AM phase does not exhibit spin susceptibility. In the presence of a nonzero magnetic field (Fig.~\ref{fig:chi005}), the PM phase is destroyed, as expected, because the magnetic field opens a gap in the chiral spin-diffuson modes. Remarkably, the magnetic field does not destroy the AM phase at low magnetic fields, and the FM order parameter $m_3$ penetrates into the AM phase (although the penetration length is short), generating the coexistence of both phases. The FM component induces a nonzero magnetic susceptibility $\chi(g,\lambda,B_3)$ (Fig.~\ref{fig:chi005}) within the AM phase (Figs.\ref{fig:gfixedB}, \ref{fig:lambdafixedB}).

\begin{figure}
\centering
\includegraphics[width=0.9\linewidth]{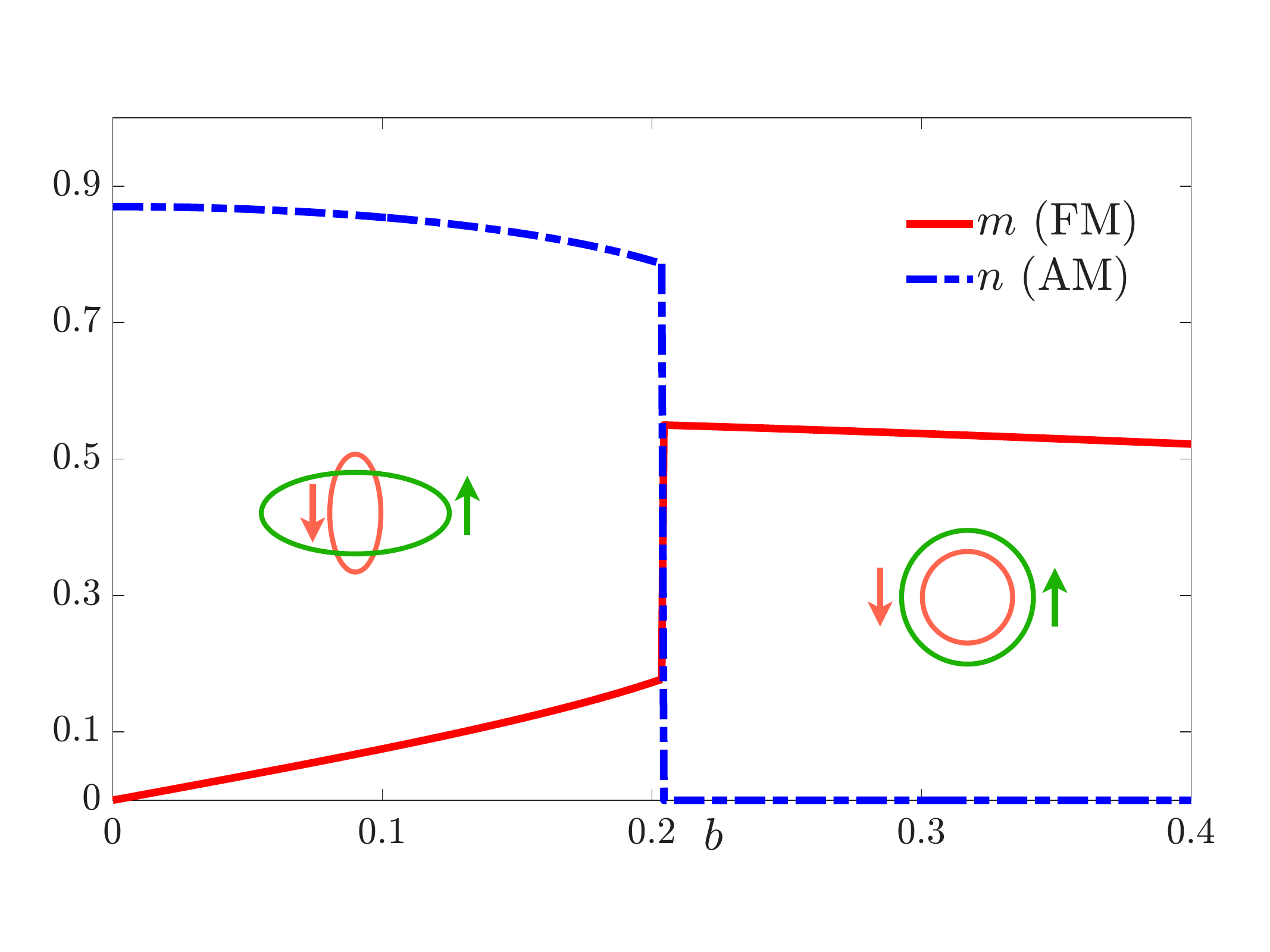}
    \caption{(Color online) Magnetization curve $m$ vs. $b$, for $\rho_R=0.4$, $g=1.5$, and $\lambda=1.8$. The transition AM--FM is first order. The inset figures represent schematically the shape of the spin-polarized Fermi surfaces on each side of the phase transition.}
    \label{fig:magnetizationcurve}
\end{figure}

In Fig.~\ref{fig:magnetizationcurve} we plot the FM (red continuous line) and AM (blue, dashed-dotted line) as a function of the dimensionless magnetic field $b$. It is interesting to note that the AM order parameter is not destroyed by the presence of an external magnetic field. In clean itinerant systems, the antiferromagnetic order parameter (the same order parameter described by $n_3$ but in the absence of staggered anisotropy $\delta$) condenses because a degeneracy between electron and hole Fermi surfaces with opposite spin is present. Adding a Zeeman term lifts the energies of bands with different spin projections, destroying this degeneracy of spin-opposed, electron-hole bands. However, in the present case, the formation of an AM phase is due to the dynamics of the spin diffuson modes, and the net effect of a Zeeman term is to change the spin precession frequency, but it simply renormalizes the spin relaxation time (see Sec.~\ref{sec:discussion}) until the AM-FM phase transition takes place at a critical value $b_c$ of the magnetic field. A feature of experimental interest is that, within the AM phase, the system displays a paramagnetic response, in which the magnetization behaves linearly with the magnetic field until $b_c$ is reached. Beyond that point, the AM order parameter becomes zero and the magnetization shows metamagnetic-like behavior. Such metamagnetic behavior should be observed in magneto-transport experiments.

\section{\label{sec:discussion}Discussion}

The first thing to mention is that we have worked in the limit of small $m_3$, $n_3$ and $\alpha_R\, k_F$ compared with the cutoff $\Lambda\sim1/\tau$ (Dyakonov--Perel approximation). This approximation has several consequences. First, the computed spin precession frequency and spin relaxation time depend only on the magnetization and the antiferromagnetic order parameter, respectively. It is expected that higher order terms in the expansion $|\bm{h}(\bm k)|\tau$ in the Green functions (\ref{eq:Greenfunction}) might introduce dependencies in the effective precession frequency with $n_3$ and $\Gamma_{\text{DP}}$ with $m_3$\cite{Burkov04}. Also, it is worth stressing that the effective potential $V_{\text{eff}}$ is computed in the one-loop approximation in the spin-diffuson modes. Is it possible to go beyond this approximation and consider how fluctuations around the corresponding magnetic order parameter (FM or AM) modify the mean field potential after integrating them out\cite{coleman73}.? These calculations go beyond the scope of the present work.

A nonzero ferromagnetic component has been experimentally observed in bulk $\text{MnTe}$ through Anomalous Hall Effect (AHE) measurements, close to the Néel temperature (AM-PM phase transition)\cite{Kluczyk2024}. A hysteretic behavior in the AHE is also reported (and attributed to the formation of Dzyaloshinskii–Moriya interactions), which is consistent with a first order phase transition scenario presented in this work. An important difference with our model is that this magnetization appears even at zero magnetic fields, while Fig.~\ref{fig:magnetizationcurve} shows zero magnetization at zero field in the AM phase. A linear dependence of the magnetization with the external magnetic field will lead to a linear transverse magnetoconductivity\cite{zyuzin21,sunko25}. The existence of a tricritical point at finite spin--orbit coupling also suggests that, in the vicinity of this point, the system is highly sensitive to perturbations such as strain or external magnetic fields, which might drive it into any of these phases, with potential consequences for experimental observations.

We have mentioned in Sec.~\ref{sec:model} that we work with the Hamiltonian (\ref{eq:kineticH}) around the $\Gamma$ point. The feature of this point is that the presence of the anisotropy parameter $\delta$ appears at order $k^4$, which is neglected, leading to the idea that we are treating an isotropic system in the absence of interactions. If we wanted to describe a system in which the Fermi level lies close to the $X$ and $Y$ points, the \emph{free} dispersion relation around these points would already be anisotropic at quadratic order, $\varepsilon(\bm k)\sim k^2_1/2m_1+k^2_2/2m_2$, with the parameters $m_1$ and $m_2$ depending on $\delta$. If we keep considering isotropic scattering described by $\Gamma_0$, it is known that the main effect is a redefinition of the diffusion coefficient $D\to D(\delta)$\cite{Wolfle1984}. The two main changes in the results presented in this work are, first, that the effective coupling constant $g=U/\pi^2D$ would now depend on $\delta$ through the diffusion constant, and second, that the coefficients of the spin-relaxation terms in Eq.~(\ref{spindiffusons}) would change because the Rashba term is modified by this momentum rescaling. The overall change will thus be that the boundaries described in Sec.~\ref{sec:phasediagram} will deform. However, we stress that the disorder analyzed in this work is invariant under $\{C_n\Vert C_2^{S}\}$ ($\text{C}_4\cdot\text{T}$ when spin--orbit is present), so it is not expected that any phase already present in the plotted phase diagrams will be destroyed in Fig.~\ref{fig:threepanels}.

Finally, we comment that the presence of the altermagnetic order parameter does not modify the localization properties of the system in the presence of spin--orbit coupling\cite{virtanen22,Hijano2024}, so the localization behavior follows standard routes\cite{Altshuler81,Finkelstein84,kirkpatrick00}. This implies that the diffusive dressing of the Coulomb interaction is the same as in the isotropic systems (the charge diffuson remains the same), leading to localization, $D\to 0$\cite{zeng25,li25}. It is important to stress that the magnetic instabilities predate the metal--insulator transition\cite{kirkpatrick00,Nayak2003}. This metal--insulator transition has been observed in disordered ultrathin samples of $\text{Ru}\text{O}_2$, originally proposed as an AM candidate\cite{Osofsky16}. Since the effective coupling constant $g=U/\pi^2D$ scales inversely with the diffusion coefficient $D$, close to the transition it blows up, implying that the system flows to the strong coupling regime of the phase diagram in Fig.~\ref{fig:threepanels} while the system becomes metallic in the AM phase at distances shorter than the localization length.

\section{Data availability}
The data supporting the figures and phase diagram in this article are not publicly available. These data are available from the author upon reasonable request.

\begin{acknowledgments}
A. C. acknowledges financial support from the Ministerio de Ciencia e Innovaci\'on through the grant PID2024-161156NB-I00. A. C. also acknowledges Severo Ochoa Centres of Excellence program through Grant CEX2024-001445-S.
\end{acknowledgments}

\appendix

\section{Computation of the effective $\bm{k}\cdot\bm{p}$ model}\label{app:kpmodel}
We will rotate the mean-field Hamiltonian operator defined in Eqs.(\ref{eq:kineticH},\ref{Hmeanfield2}) to a bonding/antibonding basis, $b^+_{\sigma}=\frac{1}{\sqrt{2}}(c^+_{A\sigma}-c^+_{B\sigma})$, $a^+_{\sigma}=\frac{1}{\sqrt{2}}(c^+_{A\sigma}+c^+_{B\sigma})$:

\begin{eqnarray}\label{eq:Happ}
H(\bm{k})&=&\begin{pmatrix}
\epsilon_0(\bm k)+\epsilon_3(\bm k) & \epsilon_1(\bm k)\\
\epsilon_1(\bm k) & \epsilon_0(\bm k)-\epsilon_3(\bm k)
\end{pmatrix},
\end{eqnarray}
where $\epsilon_0(\bm k)=t(\cos2ak_1+\cos2ak_2)+\sigma m_3$, $\epsilon_3(\bm k)=t_0(\cos ak_1+\cos ak_2)$, and $\epsilon_1(\bm k)=\Delta(\cos2ak_1-\cos2ak_2)+\sigma n_3$. The hopping amplitude $t_0$ defines the largest energy scale of this Hamiltonian operator, so we can project out the antibonding $a_{\sigma}$ states and obtain an effective Hamiltonian for the bonding states $b_{\sigma}$ after expanding around the $\Gamma$ point. The standard procedure for performing this projection is the Schrieffer--Wolff transformation. In general, we define the transformed Hamiltonian $H'$ through the transformation $H'=e^{S}H e^{-S}=e^{S}(H_0+V) e^{-S}$, where $H_0$ is the diagonal part of (\ref{eq:Happ}), and $V$ the offdiagonal part proportional to $\Delta$ and $\sigma n_3$. As usual, expanding in powers of the unknown matrix $S$, we can write,
\begin{eqnarray}\label{eq:SWtransf}
H'&=&H+[S,H]+\frac{1}{2}[S,[S,H]]+...\\\nonumber
&=&H_0+V+[S,H_0]+[S,V]+\frac{1}{2}[S,[S,H_0]]+...
\end{eqnarray}
Assuming $S$ of the order of the small parameters $\Delta/t_0$, $n_3/t_0$, we can use the ansatz, around the $\Gamma$ point, $S= b^{+}_{\sigma}(\mathcal{K}^0_{\bm k}\sigma_0+\mathcal{K}^3_{\bm k}\sigma_3 n_3)a_{\sigma}+\text{h.c.}$, with the parameters $\mathcal{K}^0_{\bm k}$ and $\mathcal{K}^3_{\bm k}$ to be fixed. Imposing the condition $V + [S, H_0] = 0$ to fix $\mathcal{K}^0_{\bm k}$ and $ \mathcal{K}^3_{\bm k}$ and introducing them in the two last terms of the expression (\ref{eq:SWtransf}), it is straightforward to get:
\begin{widetext}
\begin{eqnarray}\label{app:bondingH}
H(\bm{k})&=&\frac{1}{2}a^2(t_0-4t)\bm k^2\sigma_0+m_3\sigma_3+\frac{1}{t_0}(2\Delta a^2(k^2_1-k^2_2)\sigma_0+\sigma_3 n_3)^2\\\nonumber
&\simeq&\frac{\bm k^2}{2m_a}\sigma_0+\left(m_3+\delta\, n_3\, a^2 (k^2_1-k^2_2)\right)\sigma_3,
\end{eqnarray}
where we have trivially defined the mass $m_a$ and the dimensionless parameter $\delta=4\Delta/t_0$, which controls the staggered anisotropy. The parameter $a$ is the lattice spacing.
\section{Computation of the polarization function $\Pi_{ij}(\varphi)$}\label{app:polarization}
The  (unaveraged) polarization function $\Pi_{ij}(\omega,\bm{q},\varphi)$ introduced in the main text is given by ($\varepsilon_{\pm}=\varepsilon\pm\omega/2$ and $\bm k_{\pm}=\bm k\pm\bm q/2$), 

\begin{eqnarray}
\Pi_{ij}(\omega,\bm{q})
=
\int\frac{d^2\bm k}{(2\pi)^2}\int\!\frac{d\varepsilon}{2\pi}\,
\mathrm{Tr}\Big[
\sigma_i
G^R\Big(\varepsilon_{+},\bm{k}_{+}\Big)\,
\sigma_j
G^A\Big(\varepsilon_{-},\bm{k}_{-}\Big)
\Big]\equiv\sum_{\varphi}\Pi_{ij}(\varphi).
\label{eq:Pi_munu_app}
\end{eqnarray}
In the following treatment, the summation over momentum will be split into the summation over $\nu\,\varepsilon_{\bm k}$, and the integration over $\varphi$, which is left explicit until the end of the calculation. Integration over $\varphi$ in Eq.~(\ref{eq:Pi_munu_app}) means angular averaging over the Fermi surface. The most efficient procedure is to expand $G^{R/A}$ to the lowest nontrivial term of $\bm h(\bm{k}_{\pm})$:
For small \(|\bm h|\tau\) we expand $G^{R/A}$ as
\begin{equation}
G^{R/A}
=
G_0^{R/A}
+
G_0^{R/A}
\big(\bm h\cdot\bm\sigma\big)
G_0^{R/A}
+
G_0^{R/A}
\big(\bm h\cdot\bm\sigma\big)
G_0^{R/A}
\big(\bm h\cdot\bm\sigma\big)
G_0^{R/A}
+ \mathcal O(h^3).
\label{eq:G_expansion_smallh}
\end{equation}
\end{widetext}
The object $G_0^{R/A}(\varepsilon_{\pm},\bm k_{\pm})$ is now the retarded/advanced Green function of a standard two-dimensional electron gas:
\begin{equation}
  G^{R/A}_0(\varepsilon,\bm k)=
\frac{1}{\varepsilon - \varepsilon_{\bm k}
\pm i/(2\tau)}.
\label{eq:Greenfunction0}
\end{equation}
We can insert (\ref{eq:G_expansion_smallh}) into (\ref{eq:Pi_munu_app}) and perform the sum over spin indices, the integrations over $\varepsilon$, and $k$ term by term:
\begin{eqnarray}
\Pi_{ij}(\omega,\bm q;\varphi)
=
\Pi^{(0)}_{ij}(\omega,\bm q,\varphi)
+
\Pi^{(1)}_{ij}(\varphi)
+
\Pi^{(2)}_{ij}(\varphi).\label{app:Pi_expansion}
\end{eqnarray}
In the expansion (\ref{app:Pi_expansion}), we have retained the dependence with $\bm q$ in the zeroth term, that will be responsible for providing the diffusive structure with $\omega$ and $\bm q$. The zeroth order is the standard spinless Polarization function,
\begin{equation}
\Pi^{(0)}_{ij}
=
2\delta_{ij}
\int\!\frac{d^2\bm k}{(2\pi)^2}
\int\!\frac{d\varepsilon}{2\pi}\,
\,G_0^R(\varepsilon_+,\bm k_+)\,
\,G_0^A(\varepsilon_-,\bm k_-),
\end{equation}
where $\text{Tr}[\,\sigma_i\sigma_j]=2\delta_{ij}$ has been used. The integral over $\varepsilon$ and $\varepsilon_{\bm k}$ is standard: 

\begin{eqnarray}\label{eq:hydropole}
\Pi^{(0)}_{ij}
&\simeq&2\pi\nu\tau\delta_{ij}(1-i\omega+iv_F q\cos\varphi-\\\nonumber
&-&\tau v^2_Fq^2\cos^2\varphi).
\end{eqnarray}

Now we can easily average over $\varphi$:
\begin{eqnarray}\label{eq:hydropole2}
\Pi^{(0)}_{ij}
\simeq2\pi\nu\tau\delta_{ij}(1-i\omega\tau-Dq^2\tau),
\end{eqnarray}
with the diffusion constant $D$ defined as $D=\frac{1}{2}v^2_F\tau$. This is the standard diffuson dependence with $\omega$ and $q$.

The next term in the $\bm h$ expansion contains the trace of three Pauli matrices. Performing the trace, it is easy to see that the only nonzero term is, by using $\text{Tr}[\,\sigma_i\sigma_j\sigma_a]=2i\varepsilon_{ija}$,
\begin{eqnarray}
&&\Pi^{(1)}_{ij}(\varphi)=-i\varepsilon_{ijk}h_k(\varphi)\\\nonumber
&&\nu\int d\varepsilon_{\bm{k}}\int\frac{d\varepsilon}{2\pi}[\,(G^{R}_0(\varepsilon) )^2G^A_0(\varepsilon) -(G^{A}_0(\varepsilon) )^2G^R_0(\varepsilon) ].\label{eq:firstterm}
\end{eqnarray}
The integral can be easily performed in the $\omega=q=0$ limit to give $-2i\tau\pi\nu$ so we get the well-known result of
\begin{equation}\label{eq:finalorder1}
\Pi^{(1)}_{ij}(\varphi)=-4\pi\nu\tau\varepsilon_{ijk}h_{k}(\varphi).
\end{equation}

Performing the angular integration over $\varphi$ we get

\begin{equation}\label{eq:finalorder12}
\Pi^{(1)}_{ij}=-4\pi\nu\tau\varepsilon_{ij3}m_3.
\end{equation}
That is, the first order term in $h$ produces the precession term induced by a finite magnetization along the $\hat{z}$ direction. Both Rashba and spin nematic terms are averaged away to this order.

The next term in the expansion of $h$ is constructed from the terms in (\ref{eq:G_expansion_smallh}):
\begin{widetext}
\begin{eqnarray}
\Pi_{ij}^{(2)}&=&\int\!\frac{d^2\bm k}{(2\pi)^2}\int\frac{d\varepsilon}{2\pi}\text{Tr}\,[\sigma_iG^R_0 \,\bm h\cdot\bm \sigma\, G^R_0\,\bm h\cdot\bm \sigma\, G^R_0\sigma_j G^{A}_0]+\text{Tr}\,[\sigma_iG^R_0\sigma_j G^{A}_0 \,\bm h\cdot\bm \sigma\, G^A_0\,\bm h\cdot\bm \sigma \,G^A_0 ]\\\nonumber
&+&\int\!\frac{d^2\bm k}{(2\pi)^2} \int\frac{d\varepsilon}{2\pi}\text{Tr}\,[\sigma_iG^R_0\, \bm h\cdot\bm \sigma\, G^R_0 \sigma_j G^A_0\,\bm h\cdot\bm \sigma\,  G^{A}_0]
\end{eqnarray}

Using repeatedly the identity $\text{Tr}\,[\sigma_i\sigma_a\sigma_j\sigma_b]=2(\delta_{ia}\delta_{jb}-\delta_{ij}\delta_{ab}+\delta_{ib}\delta_{aj})$, and contracting with the symmetric product $h_ah_b$, one obtains (prior integrating over $\varphi$),
\begin{eqnarray}\label{eq:secodtermintegral}
\Pi_{ij}^{(2)}(\varphi)=\nu\int d\varepsilon_{\bm k} \int\frac{d\varepsilon}{2\pi}\Big[\delta_{ij}h^2(\varphi)[(G^{R}_0)^3 G^A_0+G^R_0(G^{A}_0)^3-(G^{R}_0)^2 (G^{A}_0)^2]+2h_i(\varphi)h_{j}(\varphi)(G^{R}_0)^2(G^{A}_0)^2\Big].
\end{eqnarray}
\end{widetext}
The integrals over $\varepsilon$ in Eq.~(\ref{eq:secodtermintegral}) can be easily performed in the $(\omega=q=0)$ limit,

\begin{subequations}
\begin{eqnarray}
&\nu&\int d\varepsilon_{\bm k}\int\,\frac{d\varepsilon}{2\pi}((G^{R}_0(\varepsilon))^3 G^A_0(\varepsilon) \\\nonumber
&+&G^R_0(\varepsilon) (G^{A}_0(\varepsilon))^3 )=-8\pi\nu\tau^2,
\end{eqnarray}
\begin{equation}
\nu\int d\varepsilon_{\bm k}\int\,\frac{d\varepsilon}{2\pi}(G^{R}_0(\varepsilon))^2 (G^{A}_0(\varepsilon))^2=4\pi\nu\tau^2,
\end{equation}
\end{subequations}
we obtain ($h^2=h^2_1+h^2_2+h^2_3$),
\begin{equation}\label{eq:finalorder2}
\Pi_{ij}^{(2)}(\varphi)=-4\pi\nu\tau^2\big[h^2(\varphi)\delta_{ij}-h_{i}(\varphi)h_{j}(\varphi)\big].
\end{equation}

The angular averaging can be readily performed in this expression. The result is that the off-diagonal terms of $\Pi^{(2)}_{ij}$ vanish. The diagonal in-plane terms read, after defining the dimensionless parameter $\lambda=\delta^2 a^4k^4_F$ which controls the anisotropy in the problem:
\begin{eqnarray}\label{averagedorder2}
\Pi^{(2)}_{ij,\perp}=-2\pi\nu\tau^2
\begin{pmatrix}
\alpha_R^2k^2_F+\lambda n^2_3 & 0 \\
0 & \alpha_R^2k^2_F+\lambda n^2_3
\end{pmatrix},
\end{eqnarray}
and $\Pi^{(2)}_{33}=-4\pi\nu\tau^2\alpha_R^2k^2_F$.
that is, the spin nematic (altermagnetic) term enters as an extra term in the spin relaxation for the $i=(x,y)$ components of the spin diffuson. The spin relaxation term for the third component of the diffuson only comes from the conventional Rashba spin--orbit coupling, as is readily seen in Eq.~(\ref{eq:finalorder2}) when we particularize $i=j=3$.

\section{\label{app:mathdetailsintegral}Mathematical details of the integral Eq.~(\ref{integralaction})}
The action term in Eq.~(\ref{integralaction}) can be handled exactly. First, we can perform the following change of variables: $u_q=Dq^2$ Also, the summation over $r=\pm 1$ is converted into the product inside the logarithm.

\begin{equation}\label{Integral}
I=\int^{\Lambda}_0 d|\omega|\int^{\Lambda}_0 du_q\ln F(|\omega|+u_q).
\end{equation}
The next step is to work out the integral over frequencies. We define $u=|\omega|+u_q$ and another variable that spans the line $|\omega|+u_q=\text{const}$, say, $v=|\omega|$, so, the inverse change is $|\omega|=v$, $u_q=u-v$, with a trivial Jacobian $\big|\text{det}\frac{\partial(|\omega|,u_q)}{\partial(u,v)}\big|=1$. The integration limits over $|\omega|$ and $u_q$ translate into the limits  $0\leq u=|\omega|+u_q\leq 2\Lambda$ and $v\in[\text{max}(0,u-\Lambda),\text{min}(\Lambda,u)]$. Since $\ln F(u)$ does not depend on $v$, the integration over $v$ simply gives the length of the $v-$interval:
\begin{equation}
\int^{\text{min}(\Lambda,u)}_{\text{max}(0,u-\Lambda)}dv\equiv L(u),
\end{equation}
with 
\begin{equation}\label{Lfunction}
L(u) =
 \begin{cases}
    u, & 0 \le u \le \Lambda,\\[2pt]
   2\Lambda - u, & \Lambda \le u \le 2\Lambda.
 \end{cases}
\end{equation}
so,
\begin{eqnarray}\label{eq:explicitintegral}
I&=&\int^{2\Lambda}_0du\,L(u)\,\ln\,F(u).
\end{eqnarray}
The infrared contribution to the effective action for ferromagnetism at $\lambda=0$ corresponds to the first term of the expression (\ref{eq:explicitintegral}).
In the present work, $F(u)=(u+\Gamma_{\text{DP}}+\lambda\, \tau\,n^2_3)+4m^2_3$, with $\Gamma_{\text{DP}}=\tau\alpha^2 k^2_F$. 

The integral is straightforward to perform. Setting $A=\Gamma_{\text{DP}}+\lambda n^2_3+4 m^2_3$, the result is
\begin{eqnarray}\label{fullexactintegral}
I&=&\frac{1}{2}(A+2\Lambda)^2\ln(A+2\Lambda)\\\nonumber
&-&(A+\Lambda)^2\ln(A+\Lambda)+\frac{1}{2}A^2\ln A-\frac{3}{2}\Lambda^2.
\end{eqnarray}


\begin{thebibliography}{52}%
\makeatletter
\providecommand \@ifxundefined [1]{%
 \@ifx{#1\undefined}
}%
\providecommand \@ifnum [1]{%
 \ifnum #1\expandafter \@firstoftwo
 \else \expandafter \@secondoftwo
 \fi
}%
\providecommand \@ifx [1]{%
 \ifx #1\expandafter \@firstoftwo
 \else \expandafter \@secondoftwo
 \fi
}%
\providecommand \natexlab [1]{#1}%
\providecommand \enquote  [1]{``#1''}%
\providecommand \bibnamefont  [1]{#1}%
\providecommand \bibfnamefont [1]{#1}%
\providecommand \citenamefont [1]{#1}%
\providecommand \href@noop [0]{\@secondoftwo}%
\providecommand \href [0]{\begingroup \@sanitize@url \@href}%
\providecommand \@href[1]{\@@startlink{#1}\@@href}%
\providecommand \@@href[1]{\endgroup#1\@@endlink}%
\providecommand \@sanitize@url [0]{\catcode `\\12\catcode `\$12\catcode
  `\&12\catcode `\#12\catcode `\^12\catcode `\_12\catcode `\%12\relax}%
\providecommand \@@startlink[1]{}%
\providecommand \@@endlink[0]{}%
\providecommand \url  [0]{\begingroup\@sanitize@url \@url }%
\providecommand \@url [1]{\endgroup\@href {#1}{\urlprefix }}%
\providecommand \urlprefix  [0]{URL }%
\providecommand \Eprint [0]{\href }%
\providecommand \doibase [0]{http://dx.doi.org/}%
\providecommand \selectlanguage [0]{\@gobble}%
\providecommand \bibinfo  [0]{\@secondoftwo}%
\providecommand \bibfield  [0]{\@secondoftwo}%
\providecommand \translation [1]{[#1]}%
\providecommand \BibitemOpen [0]{}%
\providecommand \bibitemStop [0]{}%
\providecommand \bibitemNoStop [0]{.\EOS\space}%
\providecommand \EOS [0]{\spacefactor3000\relax}%
\providecommand \BibitemShut  [1]{\csname bibitem#1\endcsname}%
\let\auto@bib@innerbib\@empty
\bibitem [{\citenamefont {\ifmmode~\check{S}\else \v{S}\fi{}mejkal}\ \emph
  {et~al.}(2022)\citenamefont {\ifmmode~\check{S}\else \v{S}\fi{}mejkal},
  \citenamefont {Sinova},\ and\ \citenamefont {Jungwirth}}]{sinova22}%
  \BibitemOpen
  \bibfield  {author} {\bibinfo {author} {\bibfnamefont {L.}~\bibnamefont
  {\ifmmode~\check{S}\else \v{S}\fi{}mejkal}}, \bibinfo {author} {\bibfnamefont
  {J.}~\bibnamefont {Sinova}}, \ and\ \bibinfo {author} {\bibfnamefont
  {T.}~\bibnamefont {Jungwirth}},\ }\href {\doibase 10.1103/PhysRevX.12.040501}
  {\bibfield  {journal} {\bibinfo  {journal} {Phys. Rev. X}\ }\textbf {\bibinfo
  {volume} {12}},\ \bibinfo {pages} {040501} (\bibinfo {year}
  {2022})}\BibitemShut {NoStop}%
\bibitem [{\citenamefont {Song}\ \emph {et~al.}(2025)\citenamefont {Song},
  \citenamefont {Bai}, \citenamefont {Zhou}, \citenamefont {Han}, \citenamefont
  {Reichlova}, \citenamefont {Dil}, \citenamefont {Liu}, \citenamefont {Chen},\
  and\ \citenamefont {Pan}}]{song25}%
  \BibitemOpen
  \bibfield  {author} {\bibinfo {author} {\bibfnamefont {C.}~\bibnamefont
  {Song}}, \bibinfo {author} {\bibfnamefont {H.}~\bibnamefont {Bai}}, \bibinfo
  {author} {\bibfnamefont {Z.}~\bibnamefont {Zhou}}, \bibinfo {author}
  {\bibfnamefont {L.}~\bibnamefont {Han}}, \bibinfo {author} {\bibfnamefont
  {H.}~\bibnamefont {Reichlova}}, \bibinfo {author} {\bibfnamefont {J.~H.}\
  \bibnamefont {Dil}}, \bibinfo {author} {\bibfnamefont {J.}~\bibnamefont
  {Liu}}, \bibinfo {author} {\bibfnamefont {X.}~\bibnamefont {Chen}}, \ and\
  \bibinfo {author} {\bibfnamefont {F.}~\bibnamefont {Pan}},\ }\href {\doibase
  10.1038/s41578-025-00779-1} {\bibfield  {journal} {\bibinfo  {journal}
  {Nature Reviews Materials}\ }\textbf {\bibinfo {volume} {10}},\ \bibinfo
  {pages} {473} (\bibinfo {year} {2025})}\BibitemShut {NoStop}%
\bibitem [{\citenamefont {Bai}\ \emph {et~al.}(2023)\citenamefont {Bai},
  \citenamefont {Zhang}, \citenamefont {Zhou}, \citenamefont {Chen},
  \citenamefont {Wan}, \citenamefont {Han}, \citenamefont {Zhu}, \citenamefont
  {Liang}, \citenamefont {Su}, \citenamefont {Han}, \citenamefont {Pan},\ and\
  \citenamefont {Song}}]{Bai23}%
  \BibitemOpen
  \bibfield  {author} {\bibinfo {author} {\bibfnamefont {H.}~\bibnamefont
  {Bai}}, \bibinfo {author} {\bibfnamefont {Y.~C.}\ \bibnamefont {Zhang}},
  \bibinfo {author} {\bibfnamefont {Y.~J.}\ \bibnamefont {Zhou}}, \bibinfo
  {author} {\bibfnamefont {P.}~\bibnamefont {Chen}}, \bibinfo {author}
  {\bibfnamefont {C.~H.}\ \bibnamefont {Wan}}, \bibinfo {author} {\bibfnamefont
  {L.}~\bibnamefont {Han}}, \bibinfo {author} {\bibfnamefont {W.~X.}\
  \bibnamefont {Zhu}}, \bibinfo {author} {\bibfnamefont {S.~X.}\ \bibnamefont
  {Liang}}, \bibinfo {author} {\bibfnamefont {Y.~C.}\ \bibnamefont {Su}},
  \bibinfo {author} {\bibfnamefont {X.~F.}\ \bibnamefont {Han}}, \bibinfo
  {author} {\bibfnamefont {F.}~\bibnamefont {Pan}}, \ and\ \bibinfo {author}
  {\bibfnamefont {C.}~\bibnamefont {Song}},\ }\href {\doibase
  10.1103/PhysRevLett.130.216701} {\bibfield  {journal} {\bibinfo  {journal}
  {Phys. Rev. Lett.}\ }\textbf {\bibinfo {volume} {130}},\ \bibinfo {pages}
  {216701} (\bibinfo {year} {2023})}\BibitemShut {NoStop}%
\bibitem [{\citenamefont {Zhang}\ \emph {et~al.}(2024)\citenamefont {Zhang},
  \citenamefont {Bai}, \citenamefont {Han}, \citenamefont {Chen}, \citenamefont
  {Zhou}, \citenamefont {Back}, \citenamefont {Pan}, \citenamefont {Wang},\
  and\ \citenamefont {Song}}]{zhang24}%
  \BibitemOpen
  \bibfield  {author} {\bibinfo {author} {\bibfnamefont {Y.}~\bibnamefont
  {Zhang}}, \bibinfo {author} {\bibfnamefont {H.}~\bibnamefont {Bai}}, \bibinfo
  {author} {\bibfnamefont {L.}~\bibnamefont {Han}}, \bibinfo {author}
  {\bibfnamefont {C.}~\bibnamefont {Chen}}, \bibinfo {author} {\bibfnamefont
  {Y.}~\bibnamefont {Zhou}}, \bibinfo {author} {\bibfnamefont {C.~H.}\
  \bibnamefont {Back}}, \bibinfo {author} {\bibfnamefont {F.}~\bibnamefont
  {Pan}}, \bibinfo {author} {\bibfnamefont {Y.}~\bibnamefont {Wang}}, \ and\
  \bibinfo {author} {\bibfnamefont {C.}~\bibnamefont {Song}},\ }\href {\doibase
  https://doi.org/10.1002/adfm.202313332} {\bibfield  {journal} {\bibinfo
  {journal} {Advanced Functional Materials}\ }\textbf {\bibinfo {volume}
  {34}},\ \bibinfo {pages} {2313332} (\bibinfo {year} {2024})}\BibitemShut
  {NoStop}%
\bibitem [{\citenamefont {Aoyama}\ and\ \citenamefont
  {Ohgushi}(2024)}]{aoyama24}%
  \BibitemOpen
  \bibfield  {author} {\bibinfo {author} {\bibfnamefont {T.}~\bibnamefont
  {Aoyama}}\ and\ \bibinfo {author} {\bibfnamefont {K.}~\bibnamefont
  {Ohgushi}},\ }\href {\doibase 10.1103/PhysRevMaterials.8.L041402} {\bibfield
  {journal} {\bibinfo  {journal} {Phys. Rev. Mater.}\ }\textbf {\bibinfo
  {volume} {8}},\ \bibinfo {pages} {L041402} (\bibinfo {year}
  {2024})}\BibitemShut {NoStop}%
\bibitem [{\citenamefont {Wu}\ \emph {et~al.}(2024)\citenamefont {Wu},
  \citenamefont {Deng}, \citenamefont {Yin}, \citenamefont {Tong},
  \citenamefont {Tian},\ and\ \citenamefont {Zhang}}]{wu24}%
  \BibitemOpen
  \bibfield  {author} {\bibinfo {author} {\bibfnamefont {Y.}~\bibnamefont
  {Wu}}, \bibinfo {author} {\bibfnamefont {L.}~\bibnamefont {Deng}}, \bibinfo
  {author} {\bibfnamefont {X.}~\bibnamefont {Yin}}, \bibinfo {author}
  {\bibfnamefont {J.}~\bibnamefont {Tong}}, \bibinfo {author} {\bibfnamefont
  {F.}~\bibnamefont {Tian}}, \ and\ \bibinfo {author} {\bibfnamefont
  {X.}~\bibnamefont {Zhang}},\ }\href {\doibase 10.1021/acs.nanolett.4c02554}
  {\bibfield  {journal} {\bibinfo  {journal} {Nano Letters}\ }\textbf {\bibinfo
  {volume} {24}},\ \bibinfo {pages} {10534} (\bibinfo {year}
  {2024})}\BibitemShut {NoStop}%
\bibitem [{\citenamefont {Zhu}\ \emph {et~al.}(2024)\citenamefont {Zhu},
  \citenamefont {Chen}, \citenamefont {Li}, \citenamefont {Qiao}, \citenamefont
  {Ma}, \citenamefont {Liu}, \citenamefont {Hu}, \citenamefont {Gao},\ and\
  \citenamefont {Ren}}]{zhu24}%
  \BibitemOpen
  \bibfield  {author} {\bibinfo {author} {\bibfnamefont {Y.}~\bibnamefont
  {Zhu}}, \bibinfo {author} {\bibfnamefont {T.}~\bibnamefont {Chen}}, \bibinfo
  {author} {\bibfnamefont {Y.}~\bibnamefont {Li}}, \bibinfo {author}
  {\bibfnamefont {L.}~\bibnamefont {Qiao}}, \bibinfo {author} {\bibfnamefont
  {X.}~\bibnamefont {Ma}}, \bibinfo {author} {\bibfnamefont {C.}~\bibnamefont
  {Liu}}, \bibinfo {author} {\bibfnamefont {T.}~\bibnamefont {Hu}}, \bibinfo
  {author} {\bibfnamefont {H.}~\bibnamefont {Gao}}, \ and\ \bibinfo {author}
  {\bibfnamefont {W.}~\bibnamefont {Ren}},\ }\href {\doibase
  10.1021/acs.nanolett.3c04330} {\bibfield  {journal} {\bibinfo  {journal}
  {Nano Letters}\ }\textbf {\bibinfo {volume} {24}},\ \bibinfo {pages} {472}
  (\bibinfo {year} {2024})}\BibitemShut {NoStop}%
\bibitem [{\citenamefont {Chakraborty}\ and\ \citenamefont
  {Black-Schaffer}(2025)}]{Schaffer25}%
  \BibitemOpen
  \bibfield  {author} {\bibinfo {author} {\bibfnamefont {D.}~\bibnamefont
  {Chakraborty}}\ and\ \bibinfo {author} {\bibfnamefont {A.~M.}\ \bibnamefont
  {Black-Schaffer}},\ }\href {\doibase 10.1103/cv8s-tk4c} {\bibfield  {journal}
  {\bibinfo  {journal} {Phys. Rev. Lett.}\ }\textbf {\bibinfo {volume} {135}},\
  \bibinfo {pages} {026001} (\bibinfo {year} {2025})}\BibitemShut {NoStop}%
\bibitem [{\citenamefont {Jiang}\ \emph {et~al.}(2025)\citenamefont {Jiang},
  \citenamefont {Liu},\ and\ \citenamefont {Wang}}]{Jiang2025}%
  \BibitemOpen
  \bibfield  {author} {\bibinfo {author} {\bibfnamefont {Y.}~\bibnamefont
  {Jiang}}, \bibinfo {author} {\bibfnamefont {H.-L.}\ \bibnamefont {Liu}}, \
  and\ \bibinfo {author} {\bibfnamefont {J.}~\bibnamefont {Wang}},\ }\href
  {\doibase 10.1088/1674-1056/add7aa} {\bibfield  {journal} {\bibinfo
  {journal} {Chinese Physics B}\ }\textbf {\bibinfo {volume} {34}},\ \bibinfo
  {pages} {107803} (\bibinfo {year} {2025})}\BibitemShut {NoStop}%
\bibitem [{\citenamefont {Alipourzadeh}\ and\ \citenamefont
  {Hajati}(2025)}]{Alipourzadeh25}%
  \BibitemOpen
  \bibfield  {author} {\bibinfo {author} {\bibfnamefont {M.}~\bibnamefont
  {Alipourzadeh}}\ and\ \bibinfo {author} {\bibfnamefont {Y.}~\bibnamefont
  {Hajati}},\ }\href {\doibase 10.1103/mj4b-2fnr} {\bibfield  {journal}
  {\bibinfo  {journal} {Phys. Rev. B}\ }\textbf {\bibinfo {volume} {111}},\
  \bibinfo {pages} {214515} (\bibinfo {year} {2025})}\BibitemShut {NoStop}%
\bibitem [{\citenamefont {Ghorashi}\ \emph {et~al.}(2024)\citenamefont
  {Ghorashi}, \citenamefont {Hughes},\ and\ \citenamefont {Cano}}]{ghorashi24}%
  \BibitemOpen
  \bibfield  {author} {\bibinfo {author} {\bibfnamefont {S.~A.~A.}\
  \bibnamefont {Ghorashi}}, \bibinfo {author} {\bibfnamefont {T.~L.}\
  \bibnamefont {Hughes}}, \ and\ \bibinfo {author} {\bibfnamefont
  {J.}~\bibnamefont {Cano}},\ }\href {\doibase 10.1103/PhysRevLett.133.106601}
  {\bibfield  {journal} {\bibinfo  {journal} {Phys. Rev. Lett.}\ }\textbf
  {\bibinfo {volume} {133}},\ \bibinfo {pages} {106601} (\bibinfo {year}
  {2024})}\BibitemShut {NoStop}%
\bibitem [{\citenamefont {Hadjipaschalis}\ \emph {et~al.}(2025)\citenamefont
  {Hadjipaschalis}, \citenamefont {Ghorashi},\ and\ \citenamefont
  {Cano}}]{Hadjipaschalis25}%
  \BibitemOpen
  \bibfield  {author} {\bibinfo {author} {\bibfnamefont {A.}~\bibnamefont
  {Hadjipaschalis}}, \bibinfo {author} {\bibfnamefont {S.~A.~A.}\ \bibnamefont
  {Ghorashi}}, \ and\ \bibinfo {author} {\bibfnamefont {J.}~\bibnamefont
  {Cano}},\ }\href {\doibase 10.1103/p79l-rty6} {\bibfield  {journal} {\bibinfo
   {journal} {Phys. Rev. B}\ }\textbf {\bibinfo {volume} {112}},\ \bibinfo
  {pages} {214430} (\bibinfo {year} {2025})}\BibitemShut {NoStop}%
\bibitem [{\citenamefont {Zhou}\ \emph {et~al.}(2025)\citenamefont {Zhou},
  \citenamefont {Cheng}, \citenamefont {Hu}, \citenamefont {Chu}, \citenamefont
  {Bai}, \citenamefont {Han}, \citenamefont {Liu}, \citenamefont {Pan},\ and\
  \citenamefont {Song}}]{Zhou25}%
  \BibitemOpen
  \bibfield  {author} {\bibinfo {author} {\bibfnamefont {Z.}~\bibnamefont
  {Zhou}}, \bibinfo {author} {\bibfnamefont {X.}~\bibnamefont {Cheng}},
  \bibinfo {author} {\bibfnamefont {M.}~\bibnamefont {Hu}}, \bibinfo {author}
  {\bibfnamefont {R.}~\bibnamefont {Chu}}, \bibinfo {author} {\bibfnamefont
  {H.}~\bibnamefont {Bai}}, \bibinfo {author} {\bibfnamefont {L.}~\bibnamefont
  {Han}}, \bibinfo {author} {\bibfnamefont {J.}~\bibnamefont {Liu}}, \bibinfo
  {author} {\bibfnamefont {F.}~\bibnamefont {Pan}}, \ and\ \bibinfo {author}
  {\bibfnamefont {C.}~\bibnamefont {Song}},\ }\href {\doibase
  10.1038/s41586-024-08436-3} {\bibfield  {journal} {\bibinfo  {journal}
  {Nature}\ }\textbf {\bibinfo {volume} {638}},\ \bibinfo {pages} {645}
  (\bibinfo {year} {2025})}\BibitemShut {NoStop}%
\bibitem [{\citenamefont {Kluczyk}\ \emph {et~al.}(2024)\citenamefont
  {Kluczyk}, \citenamefont {Gas}, \citenamefont {Grzybowski}, \citenamefont
  {Skupi\ifmmode~\acute{n}\else \'{n}\fi{}ski}, \citenamefont {Borysiewicz},
  \citenamefont {F\k{a}s}, \citenamefont {Suffczy\ifmmode~\acute{n}\else
  \'{n}\fi{}ski}, \citenamefont {Domagala}, \citenamefont {Grasza},
  \citenamefont {Mycielski}, \citenamefont {Baj}, \citenamefont {Ahn},
  \citenamefont {V\'yborn\'y}, \citenamefont {Sawicki},\ and\ \citenamefont
  {Gryglas-Borysiewicz}}]{Kluczyk2024}%
  \BibitemOpen
  \bibfield  {author} {\bibinfo {author} {\bibfnamefont {K.~P.}\ \bibnamefont
  {Kluczyk}}, \bibinfo {author} {\bibfnamefont {K.}~\bibnamefont {Gas}},
  \bibinfo {author} {\bibfnamefont {M.~J.}\ \bibnamefont {Grzybowski}},
  \bibinfo {author} {\bibfnamefont {P.}~\bibnamefont
  {Skupi\ifmmode~\acute{n}\else \'{n}\fi{}ski}}, \bibinfo {author}
  {\bibfnamefont {M.~A.}\ \bibnamefont {Borysiewicz}}, \bibinfo {author}
  {\bibfnamefont {T.}~\bibnamefont {F\k{a}s}}, \bibinfo {author} {\bibfnamefont
  {J.}~\bibnamefont {Suffczy\ifmmode~\acute{n}\else \'{n}\fi{}ski}}, \bibinfo
  {author} {\bibfnamefont {J.~Z.}\ \bibnamefont {Domagala}}, \bibinfo {author}
  {\bibfnamefont {K.}~\bibnamefont {Grasza}}, \bibinfo {author} {\bibfnamefont
  {A.}~\bibnamefont {Mycielski}}, \bibinfo {author} {\bibfnamefont
  {M.}~\bibnamefont {Baj}}, \bibinfo {author} {\bibfnamefont {K.~H.}\
  \bibnamefont {Ahn}}, \bibinfo {author} {\bibfnamefont {K.}~\bibnamefont
  {V\'yborn\'y}}, \bibinfo {author} {\bibfnamefont {M.}~\bibnamefont
  {Sawicki}}, \ and\ \bibinfo {author} {\bibfnamefont {M.}~\bibnamefont
  {Gryglas-Borysiewicz}},\ }\href {\doibase 10.1103/PhysRevB.110.155201}
  {\bibfield  {journal} {\bibinfo  {journal} {Phys. Rev. B}\ }\textbf {\bibinfo
  {volume} {110}},\ \bibinfo {pages} {155201} (\bibinfo {year}
  {2024})}\BibitemShut {NoStop}%
\bibitem [{\citenamefont {Zeng}\ \emph {et~al.}(2025)\citenamefont {Zeng},
  \citenamefont {Wang}, \citenamefont {Jiao}, \citenamefont {Shang},\ and\
  \citenamefont {Chen}}]{zeng25}%
  \BibitemOpen
  \bibfield  {author} {\bibinfo {author} {\bibfnamefont {L.}~\bibnamefont
  {Zeng}}, \bibinfo {author} {\bibfnamefont {W.}~\bibnamefont {Wang}}, \bibinfo
  {author} {\bibfnamefont {Z.}~\bibnamefont {Jiao}}, \bibinfo {author}
  {\bibfnamefont {M.}~\bibnamefont {Shang}}, \ and\ \bibinfo {author}
  {\bibfnamefont {W.}~\bibnamefont {Chen}},\ }\href {\doibase
  10.1103/sw4z-f5fl} {\bibfield  {journal} {\bibinfo  {journal} {Phys. Rev. B}\
  }\textbf {\bibinfo {volume} {112}},\ \bibinfo {pages} {054204} (\bibinfo
  {year} {2025})}\BibitemShut {NoStop}%
\bibitem [{\citenamefont {Chakraborty}\ \emph {et~al.}(2025)\citenamefont
  {Chakraborty}, \citenamefont {Schmalian},\ and\ \citenamefont
  {Fernandes}}]{Chakraborty25}%
  \BibitemOpen
  \bibfield  {author} {\bibinfo {author} {\bibfnamefont {A.~R.}\ \bibnamefont
  {Chakraborty}}, \bibinfo {author} {\bibfnamefont {J.}~\bibnamefont
  {Schmalian}}, \ and\ \bibinfo {author} {\bibfnamefont {R.~M.}\ \bibnamefont
  {Fernandes}},\ }\href {\doibase 10.1103/1vqq-9kzm} {\bibfield  {journal}
  {\bibinfo  {journal} {Phys. Rev. B}\ }\textbf {\bibinfo {volume} {112}},\
  \bibinfo {pages} {035146} (\bibinfo {year} {2025})}\BibitemShut {NoStop}%
\bibitem [{\citenamefont {Viña-Bausá}\ \emph {et~al.}(2025)\citenamefont
  {Viña-Bausá}, \citenamefont {García-Blázquez}, \citenamefont {Chourasia},
  \citenamefont {Carrasco}, \citenamefont {Expósito}, \citenamefont
  {Brihuega},\ and\ \citenamefont {Palacios}}]{palacios25}%
  \BibitemOpen
  \bibfield  {author} {\bibinfo {author} {\bibfnamefont {B.}~\bibnamefont
  {Viña-Bausá}}, \bibinfo {author} {\bibfnamefont {M.~A.}\ \bibnamefont
  {García-Blázquez}}, \bibinfo {author} {\bibfnamefont {S.}~\bibnamefont
  {Chourasia}}, \bibinfo {author} {\bibfnamefont {R.}~\bibnamefont {Carrasco}},
  \bibinfo {author} {\bibfnamefont {D.}~\bibnamefont {Expósito}}, \bibinfo
  {author} {\bibfnamefont {I.}~\bibnamefont {Brihuega}}, \ and\ \bibinfo
  {author} {\bibfnamefont {J.~J.}\ \bibnamefont {Palacios}},\ }\href
  {https://arxiv.org/abs/2501.12329} {\enquote {\bibinfo {title} {Building
  unconventional magnetic phases on graphene by h atom manipulation: From
  altermagnets to lieb ferrimagnets},}\ } (\bibinfo {year} {2025}),\ \Eprint
  {http://arxiv.org/abs/2501.12329} {arXiv:2501.12329 [cond-mat.mtrl-sci]}
  \BibitemShut {NoStop}%
\bibitem [{\citenamefont {d'Ornellas}\ \emph {et~al.}(2025)\citenamefont
  {d'Ornellas}, \citenamefont {Leeb}, \citenamefont {Grushin},\ and\
  \citenamefont {Knolle}}]{ornellas25}%
  \BibitemOpen
  \bibfield  {author} {\bibinfo {author} {\bibfnamefont {P.}~\bibnamefont
  {d'Ornellas}}, \bibinfo {author} {\bibfnamefont {V.}~\bibnamefont {Leeb}},
  \bibinfo {author} {\bibfnamefont {A.~G.}\ \bibnamefont {Grushin}}, \ and\
  \bibinfo {author} {\bibfnamefont {J.}~\bibnamefont {Knolle}},\ }\href
  {\doibase https://doi.org/10.1103/zqnk-153f} {\enquote {\bibinfo {title}
  {Altermagnetism without crystal symmetry},}\ } (\bibinfo {year} {2025}),\
  \Eprint {http://arxiv.org/abs/2504.08597} {arXiv:2504.08597
  [cond-mat.str-el]} \BibitemShut {NoStop}%
\bibitem [{\citenamefont {Chakraborty}\ \emph {et~al.}(2024)\citenamefont
  {Chakraborty}, \citenamefont {Hernández}, \citenamefont {šmejkal},\ and\
  \citenamefont {Sinova}}]{Chakraborty24}%
  \BibitemOpen
  \bibfield  {author} {\bibinfo {author} {\bibfnamefont {A.}~\bibnamefont
  {Chakraborty}}, \bibinfo {author} {\bibfnamefont {R.~G.}\ \bibnamefont
  {Hernández}}, \bibinfo {author} {\bibfnamefont {L.}~\bibnamefont
  {šmejkal}}, \ and\ \bibinfo {author} {\bibfnamefont {J.}~\bibnamefont
  {Sinova}},\ }\href {https://arxiv.org/abs/2402.00151} {\enquote {\bibinfo
  {title} {Strain induced phase transition from antiferromagnet to
  altermagnet},}\ } (\bibinfo {year} {2024}),\ \Eprint
  {http://arxiv.org/abs/2402.00151} {arXiv:2402.00151 [cond-mat.mtrl-sci]}
  \BibitemShut {NoStop}%
\bibitem [{\citenamefont {Zarzuela}\ \emph {et~al.}(2025)\citenamefont
  {Zarzuela}, \citenamefont {Jaeschke-Ubiergo}, \citenamefont {Gomonay},
  \citenamefont {\ifmmode~\check{S}\else \v{S}\fi{}mejkal},\ and\ \citenamefont
  {Sinova}}]{Zarzuela25}%
  \BibitemOpen
  \bibfield  {author} {\bibinfo {author} {\bibfnamefont {R.}~\bibnamefont
  {Zarzuela}}, \bibinfo {author} {\bibfnamefont {R.}~\bibnamefont
  {Jaeschke-Ubiergo}}, \bibinfo {author} {\bibfnamefont {O.}~\bibnamefont
  {Gomonay}}, \bibinfo {author} {\bibfnamefont {L.}~\bibnamefont
  {\ifmmode~\check{S}\else \v{S}\fi{}mejkal}}, \ and\ \bibinfo {author}
  {\bibfnamefont {J.}~\bibnamefont {Sinova}},\ }\href {\doibase
  10.1103/PhysRevB.111.064422} {\bibfield  {journal} {\bibinfo  {journal}
  {Phys. Rev. B}\ }\textbf {\bibinfo {volume} {111}},\ \bibinfo {pages}
  {064422} (\bibinfo {year} {2025})}\BibitemShut {NoStop}%
\bibitem [{\citenamefont {Sun}\ \emph {et~al.}(2025)\citenamefont {Sun},
  \citenamefont {Yang},\ and\ \citenamefont {Chen}}]{Sun25}%
  \BibitemOpen
  \bibfield  {author} {\bibinfo {author} {\bibfnamefont {Y.~J.}\ \bibnamefont
  {Sun}}, \bibinfo {author} {\bibfnamefont {F.}~\bibnamefont {Yang}}, \ and\
  \bibinfo {author} {\bibfnamefont {L.~Q.}\ \bibnamefont {Chen}},\ }\href
  {\doibase 10.1103/v12v-gl4n} {\bibfield  {journal} {\bibinfo  {journal}
  {Phys. Rev. B}\ }\textbf {\bibinfo {volume} {112}},\ \bibinfo {pages}
  {024412} (\bibinfo {year} {2025})}\BibitemShut {NoStop}%
\bibitem [{\citenamefont {Vasiakin}\ and\ \citenamefont
  {Mel'nikov}(2025)}]{Vasiakin25}%
  \BibitemOpen
  \bibfield  {author} {\bibinfo {author} {\bibfnamefont {M.~M.}\ \bibnamefont
  {Vasiakin}}\ and\ \bibinfo {author} {\bibfnamefont {A.~S.}\ \bibnamefont
  {Mel'nikov}},\ }\href {\doibase 10.1103/PhysRevB.111.L100502} {\bibfield
  {journal} {\bibinfo  {journal} {Phys. Rev. B}\ }\textbf {\bibinfo {volume}
  {111}},\ \bibinfo {pages} {L100502} (\bibinfo {year} {2025})}\BibitemShut
  {NoStop}%
\bibitem [{\citenamefont {Nayak}\ and\ \citenamefont {Yang}(2003)}]{Nayak2003}%
  \BibitemOpen
  \bibfield  {author} {\bibinfo {author} {\bibfnamefont {C.}~\bibnamefont
  {Nayak}}\ and\ \bibinfo {author} {\bibfnamefont {X.}~\bibnamefont {Yang}},\
  }\href {\doibase 10.1103/PhysRevB.68.104423} {\bibfield  {journal} {\bibinfo
  {journal} {Phys. Rev. B}\ }\textbf {\bibinfo {volume} {68}},\ \bibinfo
  {pages} {104423} (\bibinfo {year} {2003})}\BibitemShut {NoStop}%
\bibitem [{\citenamefont {Maier}\ and\ \citenamefont
  {Okamoto}(2023)}]{Maier2023}%
  \BibitemOpen
  \bibfield  {author} {\bibinfo {author} {\bibfnamefont {T.~A.}\ \bibnamefont
  {Maier}}\ and\ \bibinfo {author} {\bibfnamefont {S.}~\bibnamefont
  {Okamoto}},\ }\href {\doibase 10.1103/PhysRevB.108.L100402} {\bibfield
  {journal} {\bibinfo  {journal} {Phys. Rev. B}\ }\textbf {\bibinfo {volume}
  {108}},\ \bibinfo {pages} {L100402} (\bibinfo {year} {2023})}\BibitemShut
  {NoStop}%
\bibitem [{\citenamefont {Roig}\ \emph {et~al.}(2024)\citenamefont {Roig},
  \citenamefont {Kreisel}, \citenamefont {Yu}, \citenamefont {Andersen},\ and\
  \citenamefont {Agterberg}}]{Roig24}%
  \BibitemOpen
  \bibfield  {author} {\bibinfo {author} {\bibfnamefont {M.}~\bibnamefont
  {Roig}}, \bibinfo {author} {\bibfnamefont {A.}~\bibnamefont {Kreisel}},
  \bibinfo {author} {\bibfnamefont {Y.}~\bibnamefont {Yu}}, \bibinfo {author}
  {\bibfnamefont {B.~M.}\ \bibnamefont {Andersen}}, \ and\ \bibinfo {author}
  {\bibfnamefont {D.~F.}\ \bibnamefont {Agterberg}},\ }\href {\doibase
  10.1103/PhysRevB.110.144412} {\bibfield  {journal} {\bibinfo  {journal}
  {Phys. Rev. B}\ }\textbf {\bibinfo {volume} {110}},\ \bibinfo {pages}
  {144412} (\bibinfo {year} {2024})}\BibitemShut {NoStop}%
\bibitem [{\citenamefont {Coleman}(2015)}]{coleman2015}%
  \BibitemOpen
  \bibfield  {author} {\bibinfo {author} {\bibfnamefont {P.}~\bibnamefont
  {Coleman}},\ }\href {https://books.google.es/books?id=PDg1HQAACAAJ} {\emph
  {\bibinfo {title} {Introduction to Many-Body Physics}}}\ (\bibinfo
  {publisher} {Cambridge University Press},\ \bibinfo {year}
  {2015})\BibitemShut {NoStop}%
\bibitem [{\citenamefont {Maiani}\ and\ \citenamefont
  {Souto}(2025)}]{maiani25}%
  \BibitemOpen
  \bibfield  {author} {\bibinfo {author} {\bibfnamefont {A.}~\bibnamefont
  {Maiani}}\ and\ \bibinfo {author} {\bibfnamefont {R.~S.}\ \bibnamefont
  {Souto}},\ }\href {\doibase 10.1103/f6nc-vsnx} {\bibfield  {journal}
  {\bibinfo  {journal} {Phys. Rev. B}\ }\textbf {\bibinfo {volume} {111}},\
  \bibinfo {pages} {224506} (\bibinfo {year} {2025})}\BibitemShut {NoStop}%
\bibitem [{\citenamefont {Kirkpatrick}\ and\ \citenamefont
  {Belitz}(1996)}]{kirkpatrick96}%
  \BibitemOpen
  \bibfield  {author} {\bibinfo {author} {\bibfnamefont {T.~R.}\ \bibnamefont
  {Kirkpatrick}}\ and\ \bibinfo {author} {\bibfnamefont {D.}~\bibnamefont
  {Belitz}},\ }\href {\doibase 10.1103/PhysRevB.53.14364} {\bibfield  {journal}
  {\bibinfo  {journal} {Phys. Rev. B}\ }\textbf {\bibinfo {volume} {53}},\
  \bibinfo {pages} {14364} (\bibinfo {year} {1996})}\BibitemShut {NoStop}%
\bibitem [{\citenamefont {Chamon}\ and\ \citenamefont
  {Mucciolo}(2000)}]{Chamon2000}%
  \BibitemOpen
  \bibfield  {author} {\bibinfo {author} {\bibfnamefont {C.}~\bibnamefont
  {Chamon}}\ and\ \bibinfo {author} {\bibfnamefont {E.~R.}\ \bibnamefont
  {Mucciolo}},\ }\href {\doibase 10.1103/PhysRevLett.85.5607} {\bibfield
  {journal} {\bibinfo  {journal} {Phys. Rev. Lett.}\ }\textbf {\bibinfo
  {volume} {85}},\ \bibinfo {pages} {5607} (\bibinfo {year}
  {2000})}\BibitemShut {NoStop}%
\bibitem [{\citenamefont {Brando}\ \emph {et~al.}(2016)\citenamefont {Brando},
  \citenamefont {Belitz}, \citenamefont {Grosche},\ and\ \citenamefont
  {Kirkpatrick}}]{Belitz2016}%
  \BibitemOpen
  \bibfield  {author} {\bibinfo {author} {\bibfnamefont {M.}~\bibnamefont
  {Brando}}, \bibinfo {author} {\bibfnamefont {D.}~\bibnamefont {Belitz}},
  \bibinfo {author} {\bibfnamefont {F.~M.}\ \bibnamefont {Grosche}}, \ and\
  \bibinfo {author} {\bibfnamefont {T.~R.}\ \bibnamefont {Kirkpatrick}},\
  }\href {\doibase 10.1103/RevModPhys.88.025006} {\bibfield  {journal}
  {\bibinfo  {journal} {Rev. Mod. Phys.}\ }\textbf {\bibinfo {volume} {88}},\
  \bibinfo {pages} {025006} (\bibinfo {year} {2016})}\BibitemShut {NoStop}%
\bibitem [{Note1()}]{Note1}%
  \BibitemOpen
  \bibinfo {note} {We will limit ourselves to outline this method following
  references~\cite {kirkpatrick00,Chamon2000,Nayak2003}. We recommend to the
  reader to visit these references.}\BibitemShut {Stop}%
\bibitem [{\citenamefont {Altland}\ and\ \citenamefont
  {Zirnbauer}(1997)}]{Altland97}%
  \BibitemOpen
  \bibfield  {author} {\bibinfo {author} {\bibfnamefont {A.}~\bibnamefont
  {Altland}}\ and\ \bibinfo {author} {\bibfnamefont {M.~R.}\ \bibnamefont
  {Zirnbauer}},\ }\href {\doibase 10.1103/PhysRevB.55.1142} {\bibfield
  {journal} {\bibinfo  {journal} {Phys. Rev. B}\ }\textbf {\bibinfo {volume}
  {55}},\ \bibinfo {pages} {1142} (\bibinfo {year} {1997})}\BibitemShut
  {NoStop}%
\bibitem [{\citenamefont {Efetov}\ \emph {et~al.}(1980)\citenamefont {Efetov},
  \citenamefont {Larkin},\ and\ \citenamefont {Kheml'nitskii}}]{Efetov80}%
  \BibitemOpen
  \bibfield  {author} {\bibinfo {author} {\bibfnamefont {K.~B.}\ \bibnamefont
  {Efetov}}, \bibinfo {author} {\bibfnamefont {A.~I.}\ \bibnamefont {Larkin}},
  \ and\ \bibinfo {author} {\bibfnamefont {D.~E.}\ \bibnamefont
  {Kheml'nitskii}},\ }\href
  {http://jetp.ras.ru/cgi-bin/e/index/e/52/3/p568?a=list} {\bibfield  {journal}
  {\bibinfo  {journal} {Jour. Exp. Theor. Phys.}\ }\textbf {\bibinfo {volume}
  {52}} (\bibinfo {year} {1980})}\BibitemShut {NoStop}%
\bibitem [{\citenamefont {Altland}\ \emph {et~al.}(2006)\citenamefont
  {Altland}, \citenamefont {Simons},\ and\ \citenamefont {Simons}}]{altland06}%
  \BibitemOpen
  \bibfield  {author} {\bibinfo {author} {\bibfnamefont {A.}~\bibnamefont
  {Altland}}, \bibinfo {author} {\bibfnamefont {B.}~\bibnamefont {Simons}}, \
  and\ \bibinfo {author} {\bibfnamefont {B.}~\bibnamefont {Simons}},\ }\href
  {https://books.google.es/books?id=0KMkfAMe3JkC} {\emph {\bibinfo {title}
  {Condensed Matter Field Theory}}}\ (\bibinfo  {publisher} {Cambridge
  University Press},\ \bibinfo {year} {2006})\BibitemShut {NoStop}%
\bibitem [{\citenamefont {Lu}\ \emph {et~al.}(2025)\citenamefont {Lu},
  \citenamefont {Cao}, \citenamefont {Yuan}, \citenamefont {Coleman},\ and\
  \citenamefont {Hu}}]{Chen25}%
  \BibitemOpen
  \bibfield  {author} {\bibinfo {author} {\bibfnamefont {C.}~\bibnamefont
  {Lu}}, \bibinfo {author} {\bibfnamefont {C.}~\bibnamefont {Cao}}, \bibinfo
  {author} {\bibfnamefont {H.}~\bibnamefont {Yuan}}, \bibinfo {author}
  {\bibfnamefont {P.}~\bibnamefont {Coleman}}, \ and\ \bibinfo {author}
  {\bibfnamefont {L.-H.}\ \bibnamefont {Hu}},\ }\href
  {https://arxiv.org/abs/2510.00614} {\enquote {\bibinfo {title} {Breakdown of
  stoner ferromagnetism by intrinsic altermagnetism},}\ } (\bibinfo {year}
  {2025}),\ \Eprint {http://arxiv.org/abs/2510.00614} {arXiv:2510.00614
  [cond-mat.str-el]} \BibitemShut {NoStop}%
\bibitem [{\citenamefont {Castellani}\ \emph {et~al.}(1984)\citenamefont
  {Castellani}, \citenamefont {di~Castro}, \citenamefont {Lee}, \citenamefont
  {Ma}, \citenamefont {Sorella},\ and\ \citenamefont {Tabet}}]{Castellani84}%
  \BibitemOpen
  \bibfield  {author} {\bibinfo {author} {\bibfnamefont {C.}~\bibnamefont
  {Castellani}}, \bibinfo {author} {\bibfnamefont {C.}~\bibnamefont
  {di~Castro}}, \bibinfo {author} {\bibfnamefont {P.~A.}\ \bibnamefont {Lee}},
  \bibinfo {author} {\bibfnamefont {M.}~\bibnamefont {Ma}}, \bibinfo {author}
  {\bibfnamefont {S.}~\bibnamefont {Sorella}}, \ and\ \bibinfo {author}
  {\bibfnamefont {E.}~\bibnamefont {Tabet}},\ }\href {\doibase
  10.1103/PhysRevB.30.1596} {\bibfield  {journal} {\bibinfo  {journal} {Phys.
  Rev. B}\ }\textbf {\bibinfo {volume} {30}},\ \bibinfo {pages} {1596}
  (\bibinfo {year} {1984})}\BibitemShut {NoStop}%
\bibitem [{\citenamefont {Finkel'shtein}(1983)}]{Finkelstein82}%
  \BibitemOpen
  \bibfield  {author} {\bibinfo {author} {\bibfnamefont {A.}~\bibnamefont
  {Finkel'shtein}},\ }\href {http://jetpletters.ru/ps/0/article_23070.shtml}
  {\bibfield  {journal} {\bibinfo  {journal} {Jour. Exp. Theor. Phys.}\
  }\textbf {\bibinfo {volume} {57}},\ \bibinfo {pages} {97} (\bibinfo {year}
  {1983})}\BibitemShut {NoStop}%
\bibitem [{\citenamefont {Andreev}\ and\ \citenamefont
  {Kamenev}(1998)}]{andreev98}%
  \BibitemOpen
  \bibfield  {author} {\bibinfo {author} {\bibfnamefont {A.}~\bibnamefont
  {Andreev}}\ and\ \bibinfo {author} {\bibfnamefont {A.}~\bibnamefont
  {Kamenev}},\ }\href {\doibase 10.1103/PhysRevB.58.5149} {\bibfield  {journal}
  {\bibinfo  {journal} {Phys. Rev. B}\ }\textbf {\bibinfo {volume} {58}},\
  \bibinfo {pages} {5149} (\bibinfo {year} {1998})}\BibitemShut {NoStop}%
\bibitem [{\citenamefont {Liu}\ and\ \citenamefont {Fisher}(1973)}]{Liu73}%
  \BibitemOpen
  \bibfield  {author} {\bibinfo {author} {\bibfnamefont {K.-S.}\ \bibnamefont
  {Liu}}\ and\ \bibinfo {author} {\bibfnamefont {M.~E.}\ \bibnamefont
  {Fisher}},\ }\href {\doibase 10.1007/BF00655458} {\bibfield  {journal}
  {\bibinfo  {journal} {Journal of Low Temperature Physics}\ }\textbf {\bibinfo
  {volume} {10}},\ \bibinfo {pages} {655} (\bibinfo {year} {1973})}\BibitemShut
  {NoStop}%
\bibitem [{\citenamefont {Kosterlitz}\ \emph {et~al.}(1976)\citenamefont
  {Kosterlitz}, \citenamefont {Nelson},\ and\ \citenamefont
  {Fisher}}]{Kosterlitz76}%
  \BibitemOpen
  \bibfield  {author} {\bibinfo {author} {\bibfnamefont {J.~M.}\ \bibnamefont
  {Kosterlitz}}, \bibinfo {author} {\bibfnamefont {D.~R.}\ \bibnamefont
  {Nelson}}, \ and\ \bibinfo {author} {\bibfnamefont {M.~E.}\ \bibnamefont
  {Fisher}},\ }\href {\doibase 10.1103/PhysRevB.13.412} {\bibfield  {journal}
  {\bibinfo  {journal} {Phys. Rev. B}\ }\textbf {\bibinfo {volume} {13}},\
  \bibinfo {pages} {412} (\bibinfo {year} {1976})}\BibitemShut {NoStop}%
\bibitem [{\citenamefont {Burkov}\ and\ \citenamefont
  {Balents}(2004)}]{Burkov04}%
  \BibitemOpen
  \bibfield  {author} {\bibinfo {author} {\bibfnamefont {A.~A.}\ \bibnamefont
  {Burkov}}\ and\ \bibinfo {author} {\bibfnamefont {L.}~\bibnamefont
  {Balents}},\ }\href {\doibase 10.1103/PhysRevB.69.245312} {\bibfield
  {journal} {\bibinfo  {journal} {Phys. Rev. B}\ }\textbf {\bibinfo {volume}
  {69}},\ \bibinfo {pages} {245312} (\bibinfo {year} {2004})}\BibitemShut
  {NoStop}%
\bibitem [{\citenamefont {Coleman}\ and\ \citenamefont
  {Weinberg}(1973)}]{coleman73}%
  \BibitemOpen
  \bibfield  {author} {\bibinfo {author} {\bibfnamefont {S.}~\bibnamefont
  {Coleman}}\ and\ \bibinfo {author} {\bibfnamefont {E.}~\bibnamefont
  {Weinberg}},\ }\href {\doibase 10.1103/PhysRevD.7.1888} {\bibfield  {journal}
  {\bibinfo  {journal} {Phys. Rev. D}\ }\textbf {\bibinfo {volume} {7}},\
  \bibinfo {pages} {1888} (\bibinfo {year} {1973})}\BibitemShut {NoStop}%
\bibitem [{\citenamefont {Zyuzin}(2021)}]{zyuzin21}%
  \BibitemOpen
  \bibfield  {author} {\bibinfo {author} {\bibfnamefont {V.~A.}\ \bibnamefont
  {Zyuzin}},\ }\href {\doibase 10.1103/PhysRevB.104.L140407} {\bibfield
  {journal} {\bibinfo  {journal} {Phys. Rev. B}\ }\textbf {\bibinfo {volume}
  {104}},\ \bibinfo {pages} {L140407} (\bibinfo {year} {2021})}\BibitemShut
  {NoStop}%
\bibitem [{\citenamefont {Sunko}\ \emph {et~al.}(2025)\citenamefont {Sunko},
  \citenamefont {Liu}, \citenamefont {Vila}, \citenamefont {Na}, \citenamefont
  {Tang}, \citenamefont {Kozii}, \citenamefont {Griffin}, \citenamefont
  {Moore},\ and\ \citenamefont {Orenstein}}]{sunko25}%
  \BibitemOpen
  \bibfield  {author} {\bibinfo {author} {\bibfnamefont {V.}~\bibnamefont
  {Sunko}}, \bibinfo {author} {\bibfnamefont {C.}~\bibnamefont {Liu}}, \bibinfo
  {author} {\bibfnamefont {M.}~\bibnamefont {Vila}}, \bibinfo {author}
  {\bibfnamefont {I.}~\bibnamefont {Na}}, \bibinfo {author} {\bibfnamefont
  {Y.}~\bibnamefont {Tang}}, \bibinfo {author} {\bibfnamefont {V.}~\bibnamefont
  {Kozii}}, \bibinfo {author} {\bibfnamefont {S.~M.}\ \bibnamefont {Griffin}},
  \bibinfo {author} {\bibfnamefont {J.~E.}\ \bibnamefont {Moore}}, \ and\
  \bibinfo {author} {\bibfnamefont {J.}~\bibnamefont {Orenstein}},\ }\href
  {\doibase 10.1103/33ns-8gwj} {\bibfield  {journal} {\bibinfo  {journal}
  {Phys. Rev. B}\ }\textbf {\bibinfo {volume} {112}},\ \bibinfo {pages}
  {134407} (\bibinfo {year} {2025})}\BibitemShut {NoStop}%
\bibitem [{\citenamefont {W\"olfle}\ and\ \citenamefont
  {Bhatt}(1984)}]{Wolfle1984}%
  \BibitemOpen
  \bibfield  {author} {\bibinfo {author} {\bibfnamefont {P.}~\bibnamefont
  {W\"olfle}}\ and\ \bibinfo {author} {\bibfnamefont {R.~N.}\ \bibnamefont
  {Bhatt}},\ }\href {\doibase 10.1103/PhysRevB.30.3542} {\bibfield  {journal}
  {\bibinfo  {journal} {Phys. Rev. B}\ }\textbf {\bibinfo {volume} {30}},\
  \bibinfo {pages} {3542} (\bibinfo {year} {1984})}\BibitemShut {NoStop}%
\bibitem [{\citenamefont {Virtanen}\ \emph {et~al.}(2022)\citenamefont
  {Virtanen}, \citenamefont {Bergeret},\ and\ \citenamefont
  {Tokatly}}]{virtanen22}%
  \BibitemOpen
  \bibfield  {author} {\bibinfo {author} {\bibfnamefont {P.}~\bibnamefont
  {Virtanen}}, \bibinfo {author} {\bibfnamefont {F.~S.}\ \bibnamefont
  {Bergeret}}, \ and\ \bibinfo {author} {\bibfnamefont {I.~V.}\ \bibnamefont
  {Tokatly}},\ }\href {\doibase 10.1103/PhysRevB.105.224517} {\bibfield
  {journal} {\bibinfo  {journal} {Phys. Rev. B}\ }\textbf {\bibinfo {volume}
  {105}},\ \bibinfo {pages} {224517} (\bibinfo {year} {2022})}\BibitemShut
  {NoStop}%
\bibitem [{\citenamefont {Hijano}\ \emph {et~al.}(2024)\citenamefont {Hijano},
  \citenamefont {Ili\ifmmode~\acute{c}\else \'{c}\fi{}},\ and\ \citenamefont
  {Bergeret}}]{Hijano2024}%
  \BibitemOpen
  \bibfield  {author} {\bibinfo {author} {\bibfnamefont {A.}~\bibnamefont
  {Hijano}}, \bibinfo {author} {\bibfnamefont {S.}~\bibnamefont
  {Ili\ifmmode~\acute{c}\else \'{c}\fi{}}}, \ and\ \bibinfo {author}
  {\bibfnamefont {F.~S.}\ \bibnamefont {Bergeret}},\ }\href {\doibase
  10.1103/PhysRevResearch.6.023100} {\bibfield  {journal} {\bibinfo  {journal}
  {Phys. Rev. Res.}\ }\textbf {\bibinfo {volume} {6}},\ \bibinfo {pages}
  {023100} (\bibinfo {year} {2024})}\BibitemShut {NoStop}%
\bibitem [{\citenamefont {Al'tshuler}\ and\ \citenamefont
  {Aronov}(1981)}]{Altshuler81}%
  \BibitemOpen
  \bibfield  {author} {\bibinfo {author} {\bibfnamefont {B.~L.}\ \bibnamefont
  {Al'tshuler}}\ and\ \bibinfo {author} {\bibfnamefont {A.~G.}\ \bibnamefont
  {Aronov}},\ }\href {http://jetpletters.ru/ps/0/article_23070.shtml}
  {\bibfield  {journal} {\bibinfo  {journal} {Jour. Exp. Theor. Phys.}\
  }\textbf {\bibinfo {volume} {33}},\ \bibinfo {pages} {515} (\bibinfo {year}
  {1981})}\BibitemShut {NoStop}%
\bibitem [{\citenamefont {Finkel'stein}(1984)}]{Finkelstein84}%
  \BibitemOpen
  \bibfield  {author} {\bibinfo {author} {\bibfnamefont {A.~M.}\ \bibnamefont
  {Finkel'stein}},\ }\href {\doibase 10.1007/BF01304171} {\bibfield  {journal}
  {\bibinfo  {journal} {Zeitschrift f{\"u}r Physik B Condensed Matter}\
  }\textbf {\bibinfo {volume} {56}},\ \bibinfo {pages} {189} (\bibinfo {year}
  {1984})}\BibitemShut {NoStop}%
\bibitem [{\citenamefont {Kirkpatrick}\ and\ \citenamefont
  {Belitz}(2000)}]{kirkpatrick00}%
  \BibitemOpen
  \bibfield  {author} {\bibinfo {author} {\bibfnamefont {T.~R.}\ \bibnamefont
  {Kirkpatrick}}\ and\ \bibinfo {author} {\bibfnamefont {D.}~\bibnamefont
  {Belitz}},\ }\href {\doibase 10.1103/PhysRevB.62.966} {\bibfield  {journal}
  {\bibinfo  {journal} {Phys. Rev. B}\ }\textbf {\bibinfo {volume} {62}},\
  \bibinfo {pages} {966} (\bibinfo {year} {2000})}\BibitemShut {NoStop}%
\bibitem [{\citenamefont {Li}\ \emph {et~al.}(2025)\citenamefont {Li},
  \citenamefont {Fu}, \citenamefont {Guo}, \citenamefont {Trauzettel},\ and\
  \citenamefont {Zhang}}]{li25}%
  \BibitemOpen
  \bibfield  {author} {\bibinfo {author} {\bibfnamefont {C.-A.}\ \bibnamefont
  {Li}}, \bibinfo {author} {\bibfnamefont {B.}~\bibnamefont {Fu}}, \bibinfo
  {author} {\bibfnamefont {H.}~\bibnamefont {Guo}}, \bibinfo {author}
  {\bibfnamefont {B.}~\bibnamefont {Trauzettel}}, \ and\ \bibinfo {author}
  {\bibfnamefont {S.-B.}\ \bibnamefont {Zhang}},\ }\href
  {https://arxiv.org/abs/2507.10762} {\enquote {\bibinfo {title} {Marginal
  metals and kosterlitz-thouless type phase transition in disordered
  altermagnets},}\ } (\bibinfo {year} {2025}),\ \Eprint
  {http://arxiv.org/abs/2507.10762} {arXiv:2507.10762 [cond-mat.mes-hall]}
  \BibitemShut {NoStop}%
\bibitem [{\citenamefont {Osofsky}\ \emph {et~al.}(2016)\citenamefont
  {Osofsky}, \citenamefont {Krowne}, \citenamefont {Charipar}, \citenamefont
  {Bussmann}, \citenamefont {Chervin}, \citenamefont {Pala},\ and\
  \citenamefont {Rolison}}]{Osofsky16}%
  \BibitemOpen
  \bibfield  {author} {\bibinfo {author} {\bibfnamefont {M.~S.}\ \bibnamefont
  {Osofsky}}, \bibinfo {author} {\bibfnamefont {C.~M.}\ \bibnamefont {Krowne}},
  \bibinfo {author} {\bibfnamefont {K.~M.}\ \bibnamefont {Charipar}}, \bibinfo
  {author} {\bibfnamefont {K.}~\bibnamefont {Bussmann}}, \bibinfo {author}
  {\bibfnamefont {C.~N.}\ \bibnamefont {Chervin}}, \bibinfo {author}
  {\bibfnamefont {I.~R.}\ \bibnamefont {Pala}}, \ and\ \bibinfo {author}
  {\bibfnamefont {D.~R.}\ \bibnamefont {Rolison}},\ }\href {\doibase
  10.1038/srep21836} {\bibfield  {journal} {\bibinfo  {journal} {Scientific
  Reports}\ }\textbf {\bibinfo {volume} {6}},\ \bibinfo {pages} {21836}
  (\bibinfo {year} {2016})}\BibitemShut {NoStop}%
\end{thebibliography}
%

\end{document}